# When a Few Misfits Trigger Digital Adoption Cascades

Network Reach, Threshold Heterogeneity, and Performance-Conditioned Complex Contagion in an Agent-Based Model of Firms

**Esteve Almirall**
Esade – URL
esteve.almirall@esade.edu

**Steve Willmott**
Safe Intelligence
steve@safeintelligence.ai

**Ulises Cortés**
UPC – BSC
ulises.cortes@bsc.es

## Abstract

Why does a superior technology sometimes sweep through a population of firms and sometimes stall, even when its expected returns are an order of magnitude higher? We build an explanatory agent-based model — not a forecasting model — in which firms adopt a digital technology by imitating their successful neighbours, a fast-and-frugal social-comparison heuristic in the tradition of Gigerenzer. We contrast frequency-based (bandwagon) imitation, where a firm adopts when enough of all its neighbours are digital, with performance-conditioned imitation, where it adopts when enough of the neighbours that outperform it are digital. Defining a cascade dynamically — high final adoption reached through a rapid, accelerating takeoff rather than slow accumulation — we find, in the calibrated regime studied here, that: (i) dynamic cascades are produced by performance-conditioned imitation but not by an exogenous hazard or by frequency-based imitation; (ii) on degree-controlled networks, adoption tips sharply once the network supplies enough reach (small-world rewiring $\beta^* \approx 0.41$, 95% bootstrap CI [0.39, 0.43]), a condition jointly carried by short paths, low clustering, and high-reach nodes rather than any single graph statistic; (iii) at a fixed mean threshold, concentrating receptiveness in the lower tail can trigger cascades that homogeneous thresholds cannot; and (iv) the network position of both initial adopters and low-threshold pioneers strongly conditions cascade probability. At a realistic, human-scale observation degree the mechanism is genuine complex contagion: adoptions are triggered by reinforcement from many successful exemplars, the cascade survives an explicit requirement of two or more successful digital neighbours once ignition is given an adequate, clustered seed (the Centola–Watts critical-mass signature), and — consistent with bounded cognition and the Dunbar limit — it operates within human-scale neighbourhoods and weakens beyond them. The mechanism holds over a bounded region of parameter space and the model suggests where interventions may have leverage rather than forecasting any industry. The central, carefully bounded lesson is that broad cultural change is not required for system-wide adoption: a small minority of receptive firms, placed where the network can carry their success, is enough. In short, a few misfits can change the world.

## 1. Introduction

Digital and AI-intensive firms routinely post growth rates many times those of conventional firms. Where established companies in the S&P 500 grow at roughly 10–15% per year, digital entrants often target growth of 40% to nearly 200%, producing valuation multiples that conventional businesses cannot approach. If the advantage is so large, why is adoption so uneven — explosive in some industries and regions, glacial in others? The standard answers (adoption costs, uncertainty, organisational inertia) are real but incomplete: they explain why any individual firm might hesitate, not why hesitation sometimes resolves into a system-wide cascade and sometimes does not.

**This paper is an explanatory mechanism model, not a forecast.** We do not calibrate to a specific industry or predict adoption in any real population; we isolate a social mechanism and map the conditions under which it produces system-wide adoption. The missing ingredient, we argue, is the social process through which firms come to believe the prize is real and attainable for them. Adoption of a complex, organisationally demanding technology is not a solitary optimisation; it is a socially validated decision. Firms watch their peers, and they are moved less by the mere existence of digital adopters than by the visible success of the adopters who are outperforming them.

We formalise this as performance-conditioned imitation. A firm does not adopt because digital is globally known to be superior; it adopts when enough of the neighbours who are beating it are digital. This is a social-comparison heuristic of the kind catalogued by Gigerenzer and colleagues — imitate-the-successful — a fast-and-frugal rule that economises on information and performs well under uncertainty. (Whether this rule behaves as complex contagion in the strict, reinforcement-requiring sense is an empirical question we test directly in Section 6.5; in the sparse calibrated regime studied here it does not, so we use the more careful term throughout.) The reframing separates three levers usually conflated in discussions of digital transformation: the size and placement of the initial digital seed, the reach of the observation network, and the distribution of adoption thresholds across firms.

We ask: under what network and threshold conditions can a small minority of low-threshold firms trigger system-wide digital adoption when adoption is performance-conditioned? Carefully scoped to the calibrated regime studied here, our contributions are fourfold:

1. Mechanism contrast: dynamic cascades are produced by performance-conditioned imitation but not by an exogenous hazard or by frequency-based (bandwagon) imitation.
2. Network reach: on degree-controlled networks, cascade probability rises sharply once average path length and clustering fall together (a tipping point at small-world $\beta^* \approx 0.41$); because these structural properties are collinear, we interpret the effect as a network-reach condition rather than an isolated path-length effect.
3. Threshold heterogeneity: at a fixed mean threshold, concentrating receptiveness in the lower tail (a low-threshold minority, or higher variance) can trigger cascades that homogeneous thresholds cannot; under equal-effort matching this advantage is real but not uniformly superior to lowering the mean.

4. Placement: the network position of both initial adopters and low-threshold pioneers strongly conditions cascade probability, with hub and bridge placement far more effective than random or clustered placement.

We do not claim that thresholds, social learning, or cascades are new. The contribution is to combine a firm-level performance-conditioned imitation rule with threshold heterogeneity, network reach, and seed/pioneer placement in a transparent ABM designed to separate mechanisms that are often conflated — and to delimit, through systematic robustness checks, the parameter region and rule specifications under which the mechanism operates. The thread running through the paper is a deliberately simple proposition, which the model makes precise and bounded: a few misfits can change the world.

## 2. Related work: two logics of imitation

Our model sits at the intersection of three literatures. The first is the threshold-and-cascade tradition. Granovetter (1978) showed that collective behaviour depends on the distribution of individual thresholds; Watts (2002) showed that small shocks can trigger global cascades when a connected cluster of low-threshold nodes percolates; and Centola and Macy (2007) distinguished simple from complex contagion, where behaviours requiring social reinforcement do not spread along long ties the way information does.

The second is the economics of technology diffusion and social learning: Young (2009) on diffusion through contagion, social influence, and social learning; Conley and Udry (2010) on learning from the observed outcomes of information neighbours; and Ellison and Fudenberg (1993) on learning from others' choices. This literature explains slow diffusion of beneficial technologies but pays less attention to the cascade conditions under which a small low-threshold minority of firms tips the system.

The third is fast-and-frugal heuristics. Gigerenzer and Todd (1999) and Gigerenzer and Gaissmaier (2011) show that under uncertainty, simple social heuristics such as imitate-the-successful and imitate-the-majority can outperform information-hungry strategies. We adopt this as the micro-foundation of firm behaviour.

These literatures converge on a question with deep roots in organisational theory: why do firms imitate one another? Two answers dominate. Under frequency-based, or mimetic, imitation, firms copy what is prevalent in their reference group to gain legitimacy and reduce uncertainty, irrespective of whether the practice is paying off — the bandwagon logic of mimetic isomorphism (DiMaggio and Powell 1983) and of management fashions and fads (Abrahamson 1991), and exactly the Granovetter–Watts threshold model. Under success-based, or outcome, imitation, firms copy the practices of peers who are visibly succeeding, inferring that the practice causes the success (Haunschild and Miner 1997); this is imitate-the-successful. Our two adoption rules operationalise precisely this contrast. The empirical question is therefore not merely whether cascades occur, but which logic of imitation drives them: do firms go digital because many peers have, or because the successful peers have?

Our contribution is the specific combination: the adopters are firms; the contagion is performance-conditioned; and the decisive comparison is broad receptiveness versus a small low-threshold minority interacting with network reach. We do not claim cascades, thresholds, or social learning are new; we claim this firm-level mechanism is underexamined and policy-relevant.

## 3. The model

We summarise the model in ODD style (Grimm et al. 2020); the full ODD protocol is given in Appendix A and the replication code is available (Appendix G).

### 3.1 Purpose

To explain when digital adoption among firms becomes a cascade rather than a slow incremental process, and whether a small low-threshold minority can substitute for a population-wide shift in receptiveness. The model is explanatory: it isolates a mechanism and is evaluated by fitness for that purpose, not by empirical prediction.

### 3.2 Entities, state variables, and scales

Agents are firms on the nodes of a fixed undirected observation network. Each firm has a binary digital status, a valuation, an adoption threshold, and an age (Appendix A, Table A1). Time is discrete; one period is a strategic observation-and-adjustment cycle. The baseline is N = 1,000 firms over T = 100 periods.

### 3.3 Process overview and scheduling

Each period: (1) every firm draws a return and updates its valuation; (2) firms below a viability floor exit and are replaced by traditional entrants, holding N constant; (3) each non-digital firm evaluates adoption under the experimental rule; (4) adoption decisions are computed from a frozen snapshot and committed simultaneously (synchronous update; an asynchronous variant is tested in robustness).

### 3.4 Submodel: valuation dynamics

Valuation evolves multiplicatively:

$$v_i(t+1) = v_i(t)(1 + r_i(t)) \quad (1)$$

$$r_i(t) \sim N\left(\mu_{d_i}, \sigma_{d_i}\right) \quad (2)$$

with $(\mu_T, \sigma_T) = (0.15, 0.10)$ for traditional firms and $(\mu_D, \sigma_D) = (1.5, 1.0)$ for digital firms: a tenfold mean premium with high variance, so digital success is uncertain. Returns are truncated below at −0.99; a firm with valuation below 1 exits and is replaced. Taxation and premium depreciation are excluded from the baseline.

### 3.5 Submodel: adoption rules — frequency-based and performance-conditioned imitation

Each period a non-digital firm may switch by imitating the firms it observes. Both imitation logics compare a digital exposure share to the firm's threshold $\theta_i$ (a fraction, so the rule is degree-invariant); they differ only in which neighbours count. Let $N_i$ be the neighbours of i and let the better-performing set be

$$B_i(t) = \{ j \in N_i : v_j(t) > v_i(t) \} \quad (3)$$

**Performance-conditioned (success-based) imitation** — copy the winners; exposure is the digital share among better-performing neighbours only:

$$\varphi_i^S(t) = \frac{1}{|B_i(t)|} \sum_{j \in B_i(t)} d_j(t) \quad (4)$$

If the firm has no better-performing neighbour it does not adopt; this is the operative candidate condition. The fractional rule is performance-conditioned complex contagion: a firm adopts when a sufficient share of its successful peers are digital, which at realistic neighbourhood sizes requires reinforcement from many exemplars. Whether the rule behaves as genuine reinforcement-based complex contagion or degenerates to a single-exemplar trigger is an empirical question about neighbourhood size, which we test directly: in a deliberately sparse regime (mean degree 4) the better-performing set is so small that one digital exemplar can clear the threshold, whereas at realistic, human-scale degree (Section 6.6, Experiments K and L) adoptions are triggered by many reinforcing exemplars and the cascade survives an explicit minimum-count requirement once ignition is adequate. We therefore report both regimes and are explicit about which one supports the complex-contagion reading.

**Frequency-based (bandwagon) imitation** — copy the crowd; exposure is the digital share among all neighbours, with no performance conditioning:

$$\varphi_i^F(t) = \frac{1}{|N_i|} \sum_{j \in N_i} d_j(t) \quad (5)$$

Under either logic the firm adopts when its exposure reaches its threshold, $\phi_i(t) \geq \theta_i$. The exogenous-hazard null model instead has each non-digital firm adopt with a fixed per-period probability $\lambda$, independent of neighbours. Digital status is absorbing in the baseline. Robustness checks (Sections 6.5–6.6, Appendices E and G) examine reinforcement variants that require at least two or three successful digital neighbours, an all-neighbour denominator, an absolute-count rule, a non-absorbing (abandonment) variant, and alternative performance benchmarks (own valuation, neighbourhood mean, median, or 75th percentile).

### 3.6 Submodel: organizational receptiveness (the threshold distribution)

The threshold $\theta_i$ represents a firm's organizational receptiveness to digital adoption — which may reflect risk tolerance, absorptive capacity, managerial attention, capital constraints, or culture. We reserve the word "culture" for interpretation and use "receptiveness" and "threshold distribution" in the analysis. We compare a homogeneous baseline (all firms share $\bar{\theta}$) with three mean-matched treatments: lower-mean (the whole population more receptive); higher-variance (same mean, wider spread); and a mixture in which a share $\pi$ of low-threshold pioneers ("misfits") receive θ_low while the rest are set higher so the mean is unchanged:

$$\theta_i \sim \pi \cdot N(\mu_P, \sigma) + (1 - \pi) \cdot N(\mu_H, \sigma) \quad (6)$$

### 3.7 Submodel: networks (degree-controlled)

Firms observe only network neighbours, at a common mean degree k. The small-world family uses a degree-conserving Watts–Strogatz rewiring generator (rewiring existing edges rather than adding shortcuts), so realised mean degree is held constant as the rewiring parameter β varies; this corrects a confound in earlier versions where shortcut-adding generators inflated degree with β. The scale-free family is the Barabási–Albert generator (labelled degree-varying, since degree scales with the attachment parameter m), and we add a degree-matched Erdős–Rényi control. Every realised network records mean degree, clustering, degree variance, average path length, and giant-component fraction.

### 3.8 Initial seeds versus low-threshold pioneers

We keep two constructs strictly separate. Initial digital seeds are firms that are digital at t = 0. Low-threshold pioneers (misfits) are firms with unusually low θ; they are not necessarily digital at t = 0. Experiment D varies seed placement; Experiment E crosses seed placement with pioneer placement. Placement strategies — random, high-degree (hub), bridge (high-betweenness), combined hub/bridge, and clustered (a connected community) — are defined algorithmically in Appendix A.

### 3.9 Exit and replacement dynamics

When a firm's valuation falls below the viability floor (1 unit) it exits and is replaced at the same node, so N is constant. A replacement entrant resets valuation to v0 and age to 0, draws a fresh threshold from the run's treatment distribution, and is placed before the adoption stage of the same period; because adoption is synchronous, an entrant can therefore both adopt and be observed by neighbours in the period it enters.

**A correction and an ablation.** An audit of the implementation revealed that the production code used to generate Experiments A–F set a replacement entrant to digital with probability equal to the initial seed fraction (the rule we now label *bernoulli_seed_probability*), i.e. it applied a small amount of continuing exogenous digital reseeding. This differs from the natural baseline stated in the model — that replacement entrants are traditional (the *traditional_only* rule). We expose both rules explicitly, set *traditional_only* as the model baseline, and ablate the two against each other in Experiment I (Section 6.5, Appendix I). The two rules give materially the same cascade probabilities — for example, in the Experiment A focal cell the performance-conditioned cascade probability is 1.00 under traditional-only entrants and 0.99 under Bernoulli entrants — so the headline results are not artefacts of continuing reseeding. Each results table notes which rule it used; the production A–F figures use *bernoulli_seed_probability*, and Experiment I confirms robustness to the alternative.

## 4. Outcome measures and the cascade definition

Let aggregate adoption be the digital share:

$$A(t) = \frac{1}{N}\sum_{i=1}^{N} d_i(t) \quad (7)$$

with times to ten, fifty, and ninety per cent adoption $T_{10}$, $T_{50}$, $T_{90}$ defined as the first periods crossing those levels. Crucially, we define a cascade dynamically, so that slow accumulation is not mistaken for endogenous takeoff. A run is a cascade when all three hold:

$$A(T) \geq 0.5 \quad \wedge \quad T_{50} - T_{10} \leq W \quad \wedge \quad max_t \, \Delta A(t) \geq s_{min} \quad (8)$$

with W = 45 periods and s_min = 0.02 per period in the baseline (a stricter A(T) ≥ 0.7 variant and the legacy final-adoption-only flags are retained for transparency). The cascade probability is the average of this indicator over R replications,

$$P_{cascade} = \frac{1}{R}\sum_{r=1}^{R} C_r \quad (9)$$

and every cascade probability is reported with a Wilson 95% confidence interval. This dynamic definition removes the contradiction whereby an exogenous hazard that merely accumulates to 50% would be counted as a cascade. The window W and slope s_min are necessarily conventional; Section 6.6 (Appendix H) shows the qualitative conclusions are stable across a grid of W, s_min, and final-adoption thresholds, and that the hazard and frequency-based rules are never reclassified as cascades.

## 5. Experimental design

Ten experiments isolate the mechanism, its dependencies, and its robustness (Table 1). Experiments A–F use performance-conditioned imitation (except A, which compares the three rules) and constitute the main analysis; Experiments G–J are the robustness and scoping programme added in revision. Baseline parameters are in Table 2; the regime ($\bar{\theta}$ = 0.70, seed 0.3%, k = 4) was located by calibration at the edge of the cascade transition so that treatment contrasts are visible rather than saturated. The main results are full-mode production runs (N = 1,000) using an adaptive replication design — up to 1,000 replications for focal cells and 250 near phase boundaries; the revision experiments use R = 250 (G, I) or R = 100 (J) per cell. Wilson 95% confidence intervals accompany every cascade probability.

**Table 1.** *The ten experiments (A–F main analysis; G–J robustness and scoping).*

| Exp | Question | Design |
|---|---|---|
| **A** | Which logic of imitation drives the cascade? | Compare exogenous hazard, frequency-based, and performance-conditioned imitation on a small-world β = 1.0 network with a 1% seed and π = 0.10, under the dynamic cascade metric. |
| **B** | How does topology gate it? | Sweep small-world β (degree-conserving) and scale-free m at fixed degree; add a degree-matched Erdős–Rényi control; report structural metrics; regress cascade on them. |

| **C** | Receptiveness: tail vs mean? | At fixed mean $\bar{\theta}$, compare homogeneous, lower-mean, higher-variance, and mixture ($\pi$ = 1/5/10%) on a focal $\beta$ = 0.4 network; plus an equal-effort version (Appendix D). |
|---|---|---|
| **D** | Does seed position matter? | Place initial digital seeds at random / on hubs / on bridges / clustered, on small-world and scale-free networks. |
| **E** | Seeds vs pioneers? | 2×2: cross initial-seed placement (random vs hub/bridge) with low-threshold-pioneer placement (random vs hub/bridge) at matched counts. |
| **F** | How robust is it to parameters? | Sweep digital premium, return variance and distribution, benchmark definition, update schedule, seed size, threshold mean, and absorbing vs non-absorbing adoption (Appendix E). |
| **G** | Is it complex contagion? | Stricter adoption rules: require ≥2 or ≥3 successful digital neighbours; all-neighbour denominator; absolute counts (Section 6.5, Appendix G). |
| **H** | Does the cascade metric matter? | Reclassify existing trajectories over final cutoff {0.5, 0.7}, W {30, 45, 60}, s_min {0.01–0.03} (Section 6.6, Appendix H). |
| **I** | Do replacement entrants matter? | Ablate traditional-only vs Bernoulli-seed replacement entrants across mechanism, transition, and placement cells (Section 6.5, Appendix I). |
| **J** | How wide is the cascade region? | Parameter-space maps: $\bar{\theta}$ × seed fraction, premium × variance, $\beta$ × misfit share (Section 6.7). |

**Table 2.** *Baseline parameters.*

| Parameter | Value | Note |
|---|---|---|
| Firms N | 1,000 | Constant; bankrupt firms replaced |
| Horizon T | 100 periods | Early stopping |
| Traditional return | N(0.15, 0.10) | 15% mean |
| Digital return | N(1.5, 1.0) | 10× mean, uncertain |
| Initial digital seed | 0.3% (Exp A/C/D vary) | Distinct from pioneers |
| Mean degree k | 4 | Held fixed across small-world $\beta$ |
| Threshold mean $\bar{\theta}$ | 0.70 | Edge of cascade regime (calibrated) |
| Cascade W, s_min | 45 periods, 0.02 | Dynamic cascade thresholds |

## 6. Results

### 6.1 Which logic of imitation produces cascades?

Experiment A varies only the logic of imitation. Under the dynamic cascade definition, in the calibrated regime studied here, the contrast is clear (Figure 1, Table 3). The exogenous hazard reaches a median final adoption of 0.50 but its cascade probability is 0.00: it accumulates slowly (median 10→50% transition of 83 periods, maximum slope 0.015) and never takes off. Frequency-based, bandwagon imitation does not spread at all (median final adoption 0.012, cascade probability 0.00). Performance-conditioned imitation produces dynamic cascades (median final adoption 0.90; cascade probability 0.99, 95% CI [0.98, 0.99]; transition 19 periods; maximum slope 0.031). In this model, digital transformation does not spread because adoption becomes common; it spreads because the firms that are visibly winning have adopted. The strength of the reinforcement underlying this contagion depends on neighbourhood size: Section 6.6 (Experiments K and L) shows that at a realistic, human-scale observation degree the same rule is genuine reinforcement-based complex contagion — adoptions are triggered by many successful exemplars, not one.

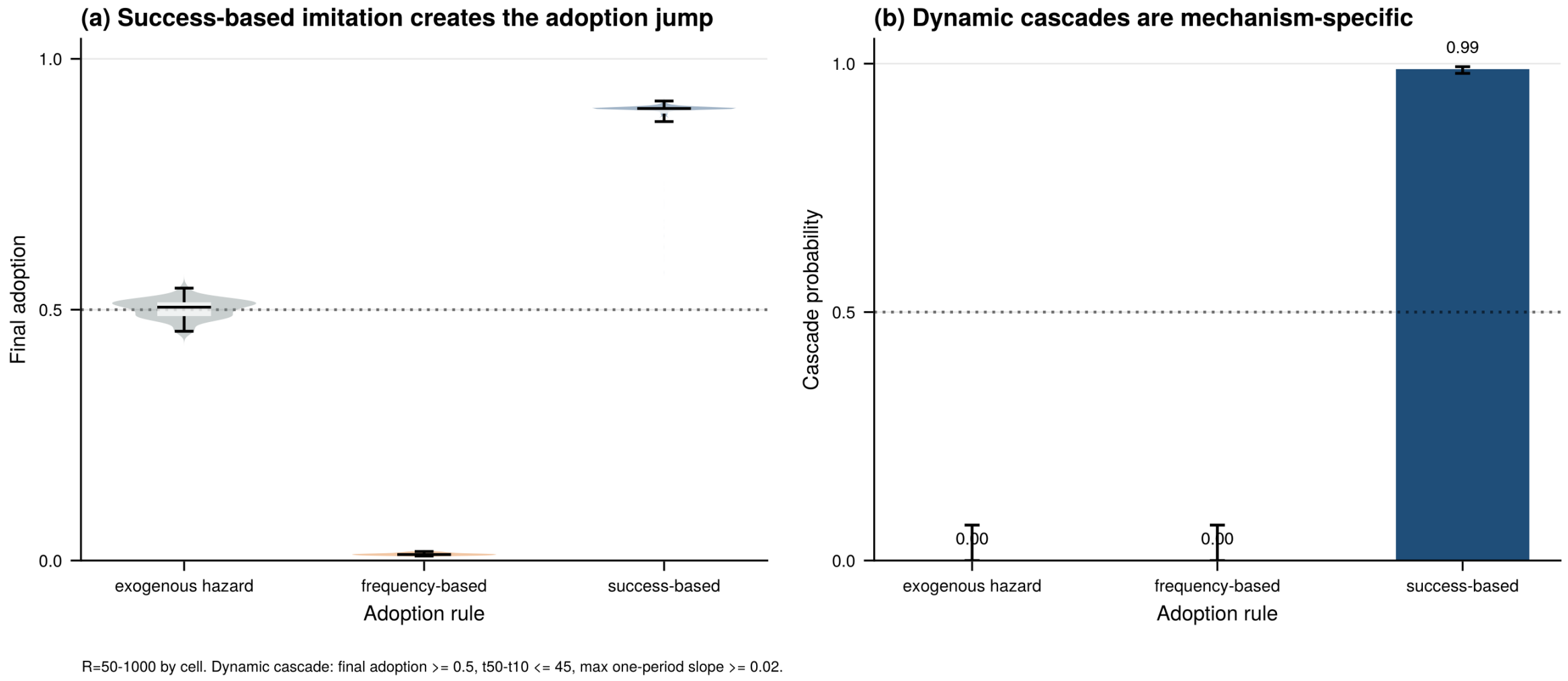

**Figure 1.** *Adoption by imitation logic under the dynamic cascade metric (small-world β = 1.0, π = 0.10, 1% seed; N = 1,000; R up to 1,000). (a) Final adoption; (b) cascade probability with Wilson 95% CIs. Plot axis labels use the rule names hazard / simple / complex.*

**Table 3.** *Experiment A (dynamic metric; medians; Wilson 95% CI on cascade probability).*

| Imitation logic | Final adopt. | Cascade prob. [CI] | Transition $T_{50}$–$T_{10}$ |
|---|---|---|---|
| Exogenous hazard | 0.50 | 0.00 [0.00, 0.07] | 83 |
| Frequency-based (simple) | 0.012 | 0.00 [0.00, 0.07] | — |
| Success-based (complex) | 0.90 | 0.99 [0.98, 0.99] | 19 |

## 6.2 A sharp tipping point gated by network reach

Experiment B sweeps topology with degree held constant. The small-world realised mean degree is fixed at 4.0 across all β, so the effect is purely structural. Cascade probability is essentially zero for clustered, long-path networks (β ≤ 0.2; average path length 30.1 at β = 0.01, 8.8 at β = 0.1, 6.9 at β = 0.2) and then rises sharply — 0.48 at β = 0.4 (path length 5.8, 95% CI [0.45, 0.51]), 0.85 at β = 0.7 (path length 5.4), and 0.94 at β = 1.0. Bootstrapping the sweep places the tipping point at β* ≈ 0.41 with a tight 95% CI of [0.39, 0.43] — higher than the β ≈ 0.2 of the earlier degree-inflating generator, confirming that part of the old effect was a degree artifact. Scale-free networks cascade even at low attachment (m = 2, cascade probability 0.58; m ≥ 3, ≈ 1.0), and a degree-matched Erdős–Rényi control at degree 4 cascades in 0.92–0.98 of runs precisely because its paths are short and its clustering low.

What structural property controls the cascade? The clean causal evidence is the degree-controlled small-world sweep itself: holding degree fixed, cascade probability jumps from near zero to near one over a narrow β range as both average path length and clustering fall together. A pooled cross-family logistic regression (Appendix C) confirms that network structure strongly predicts the cascade, but the structural covariates — average path length, clustering, and mean degree — are highly collinear across our network families, so the regression cannot cleanly attribute the effect to a single statistic; in the full sample lower clustering and higher degree carry the strongest signs while average path length is not separately identified. We therefore describe the controlling factor as network reach — short paths, low clustering, and hubs acting together — rather than any one metric. The substantive point is unchanged: sufficient reach is the necessary condition for the cascade.

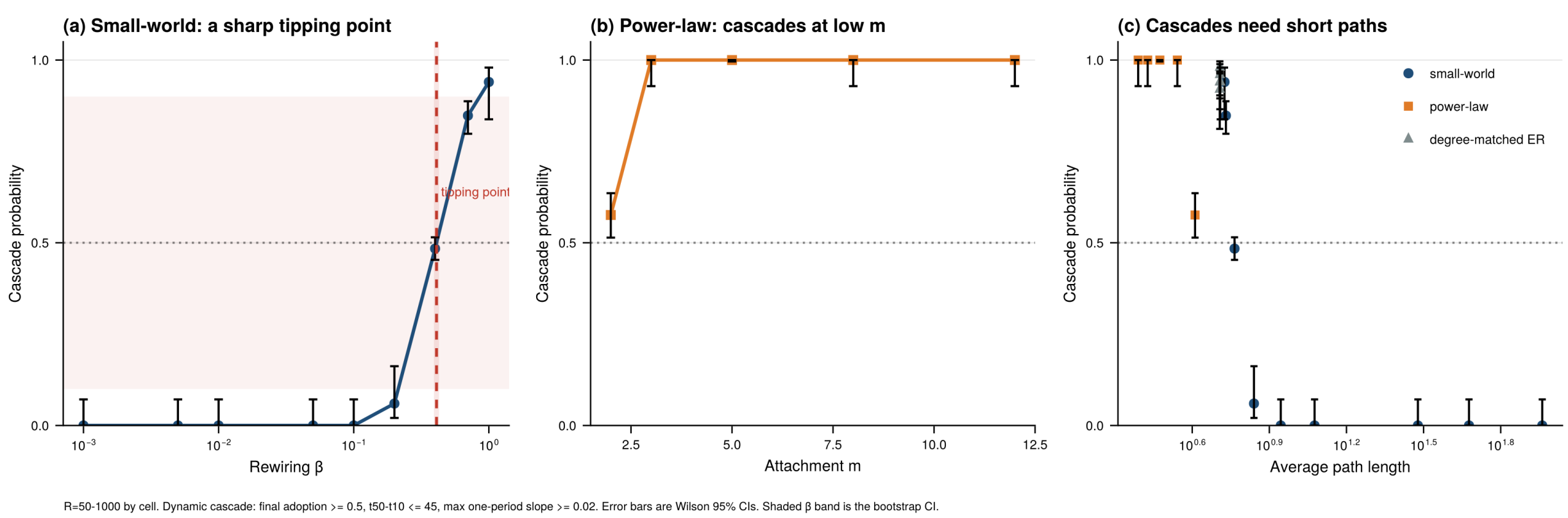

**Figure 2.** *Topology gates the cascade at fixed degree (N = 1,000; Wilson 95% CIs). (a) Small-world tipping point β* ≈ 0.41 (dashed line; shaded bootstrap CI); (b) scale-free cascades at low m; (c) pooled across families, including the degree-matched Erdős–Rényi control, cascades require short paths.*

## 6.3 Receptiveness: lower-tail heterogeneity versus the mean

Experiment C fixes the network at the transition (β = 0.4) and the mean threshold at $\bar{\theta} = 0.70$, and varies the shape of the receptiveness distribution (Figure 3, Table 4). A modest lowering of the mean helps only a little (cascade probability 0.20 → 0.36). Widening the tail transforms the system: higher variance at the same mean reaches 0.99, and a mixture with a small low-threshold minority rises from 0.38 at a 1% misfit share to 0.86 at 5% and 0.96 at 10% — even though the misfits are exactly offset by more conservative firms so the mean is unchanged. The adoption trajectories (Figure 4) show homogeneous and lower-mean societies near the boundary while variance and mixture societies climb past it.

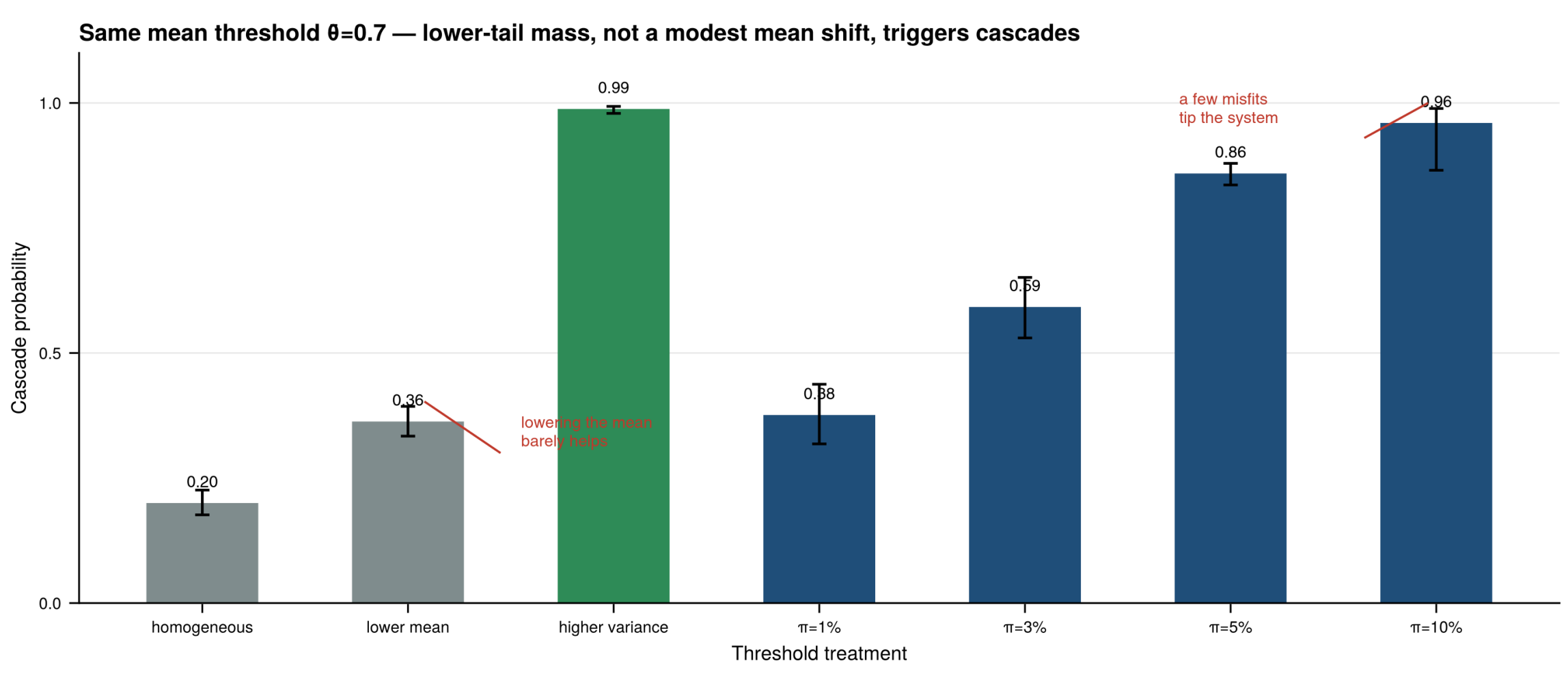

**Figure 3.** *At a fixed mean threshold ($\beta = 0.4$, $\bar{\vartheta} = 0.70$; $N = 1{,}000$), widening the lower tail — higher variance or a small misfit minority — raises cascade probability far more than a modest lowering of the mean. Wilson 95% CIs.*

**Table 4.** *Experiment C at fixed mean $\bar{\vartheta} = 0.70$, $\beta = 0.4$ (cascade probability, Wilson 95% CI).*

| Receptiveness treatment | Cascade prob. [CI] | n |
|---|---|---|
| Homogeneous | 0.20 [0.17, 0.22] | 1000 |
| Lower mean (modest) | 0.36 [0.33, 0.39] | 1000 |
| Higher variance (same mean) | 0.99 [0.97, 0.99] | 1000 |
| Mixture, π = 5% misfits | 0.86 [0.83, 0.87] | 1000 |
| Mixture, π = 10% misfits | 0.96 [0.86, 0.98] | 50 |

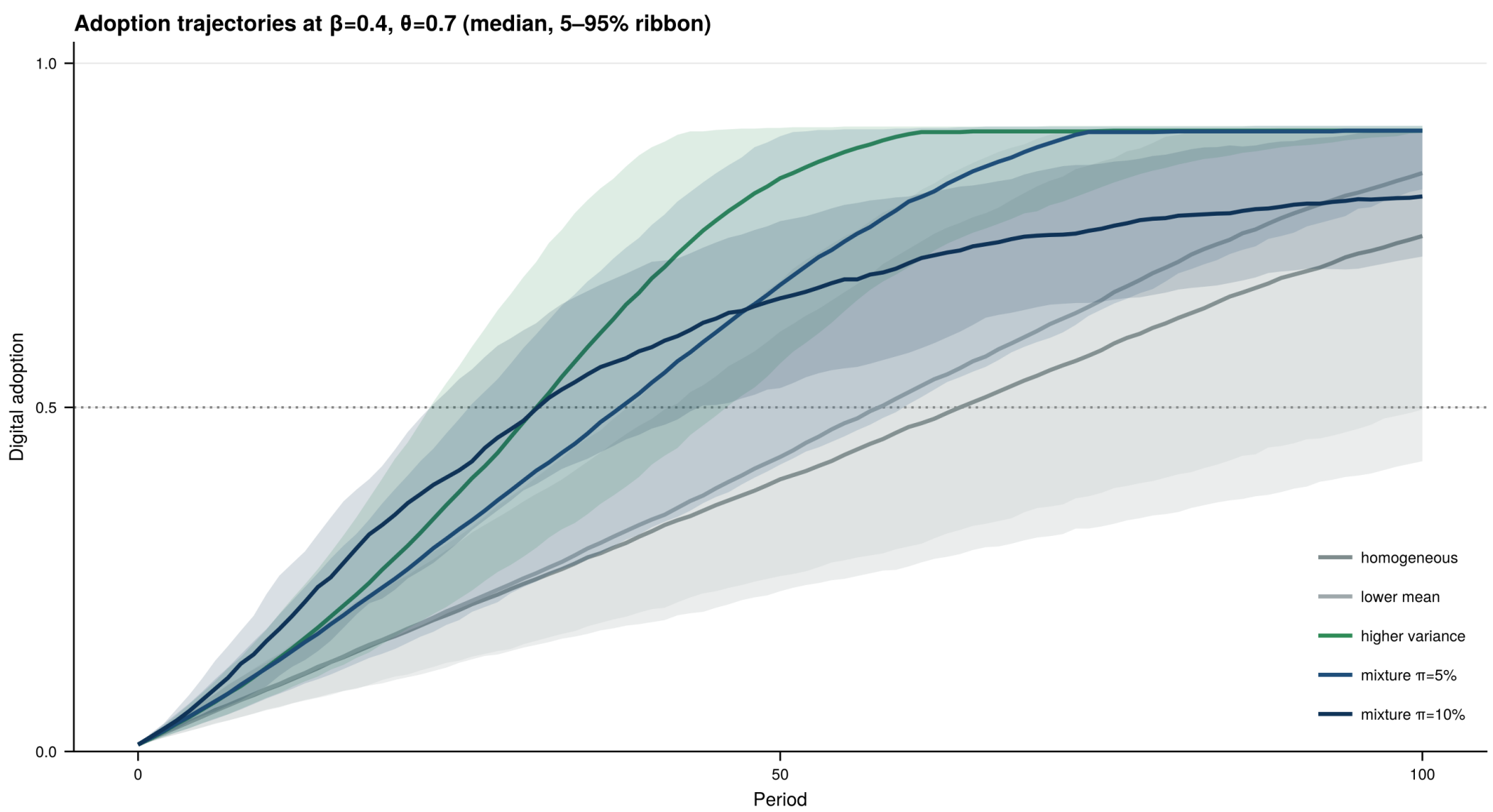

**Figure 4.** *Adoption trajectories by treatment ($\beta$ = 0.4, $\bar{\vartheta}$ = 0.70; N = 1,000; median with 5–95% ribbon).*

**Is the comparison fair? An equal-effort test.** Because the lower-mean condition above is a deliberately modest shift while the variance/mixture conditions reshape the tail substantially, we re-ran the comparison matching the treatments on a common budget — equal L1 distance of the threshold distribution from baseline, and equal expected mass below a critical threshold (Appendix D, Figure D1). Under equal effort the picture is more nuanced: at a small L1 budget (0.02) the mixture leads narrowly (0.40 vs 0.34 for higher variance and 0.37 for lower mean), but at a larger L1 budget (0.05) a matched lowering of the mean overtakes it (lower-mean 1.00 vs mixture 0.74), and under the mass-below-threshold budget lowering the mean or raising variance reaches 1.00 while the mixture trails (0.57–0.87). The honest, defensible claim is therefore: at a fixed mean, concentrating receptiveness in the lower tail is at least as effective as homogeneous thresholds and can cascade where homogeneity cannot, but it is not uniformly superior to an equal-effort reduction of the mean. The robust, decisive levers are network reach and placement (Sections 6.2 and 6.4).

The interaction between the misfit share and connectivity is shown in Figure 5: misfits ignite cascades only where the network already supplies reach; at short path lengths, increasing the misfit share moves the system from no cascade to near-certain cascade.

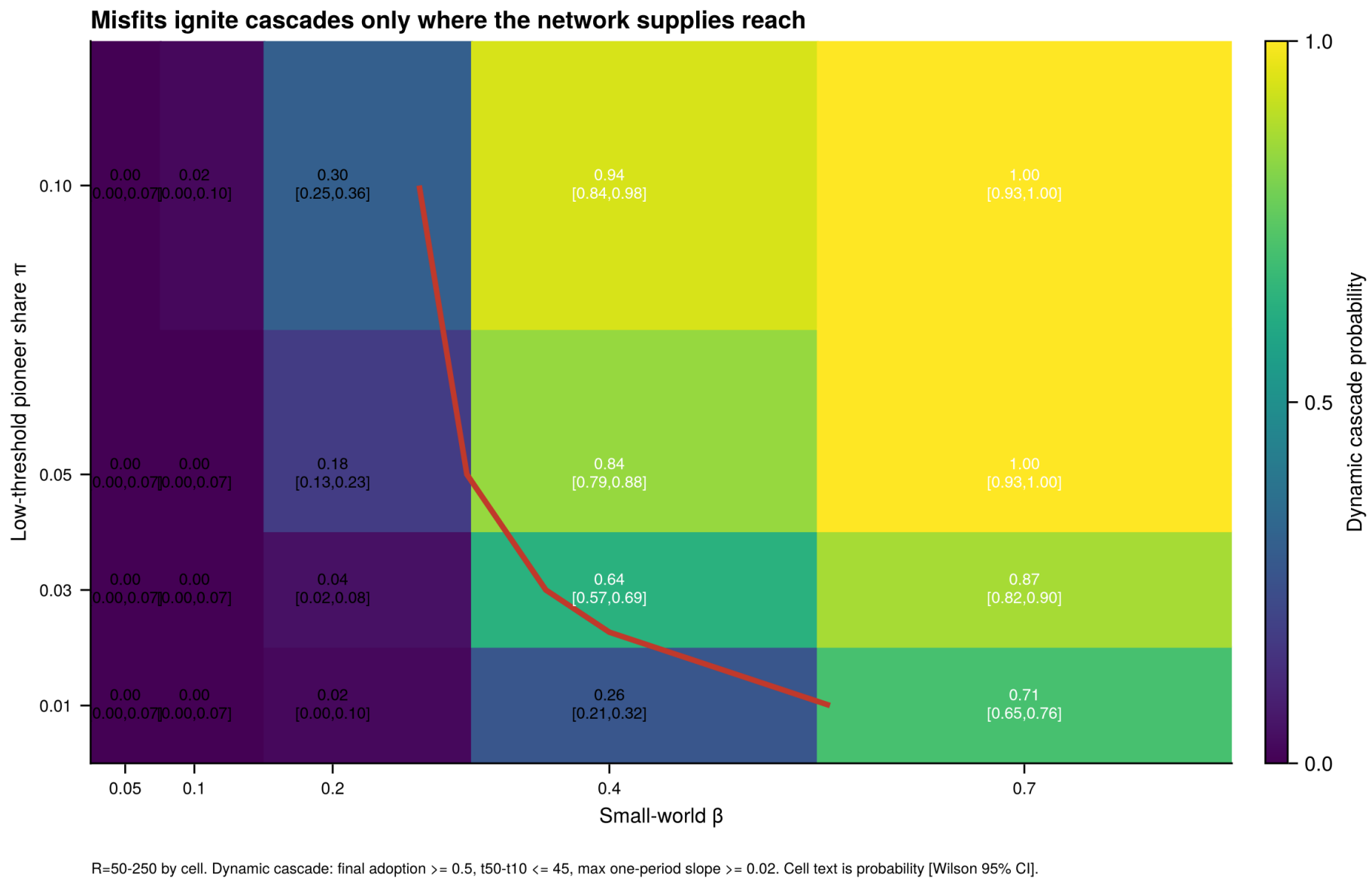

**Figure 5.** *Cascade probability over misfit share (π) and connectivity (N = 1,000). Misfits matter only where the network gives them reach.*

## 6.4 Position is decisive: seeds and pioneers

Experiment D varies where the same initial digital seeds are placed (Figure 6, Table 5). On a small-world network at the transition, hub and bridge placement raise cascade probability to 0.86–0.88, random placement reaches 0.47, and clustering the seeds in one community suppresses it to 0.16. On scale-free networks hub/bridge placement reaches 1.00 while random placement is only 0.33. Experiment E then separates the two constructs in a 2×2 (Table 6): placing low-threshold pioneers on hubs/bridges raises cascade probability to 1.00 regardless of seed placement, while random pioneers with random seeds reach only 0.51; placing initial seeds well helps too (0.92 with random pioneers). Both the position of the initial exemplars and the position of the receptive pioneers matter, and pioneer placement on bridges is especially powerful — it also roughly halves the time to half-adoption (median $T_{50}$ from 54 to 24 periods). The policy claim sharpens from "a few misfits" to "a few misfits in the right network positions."

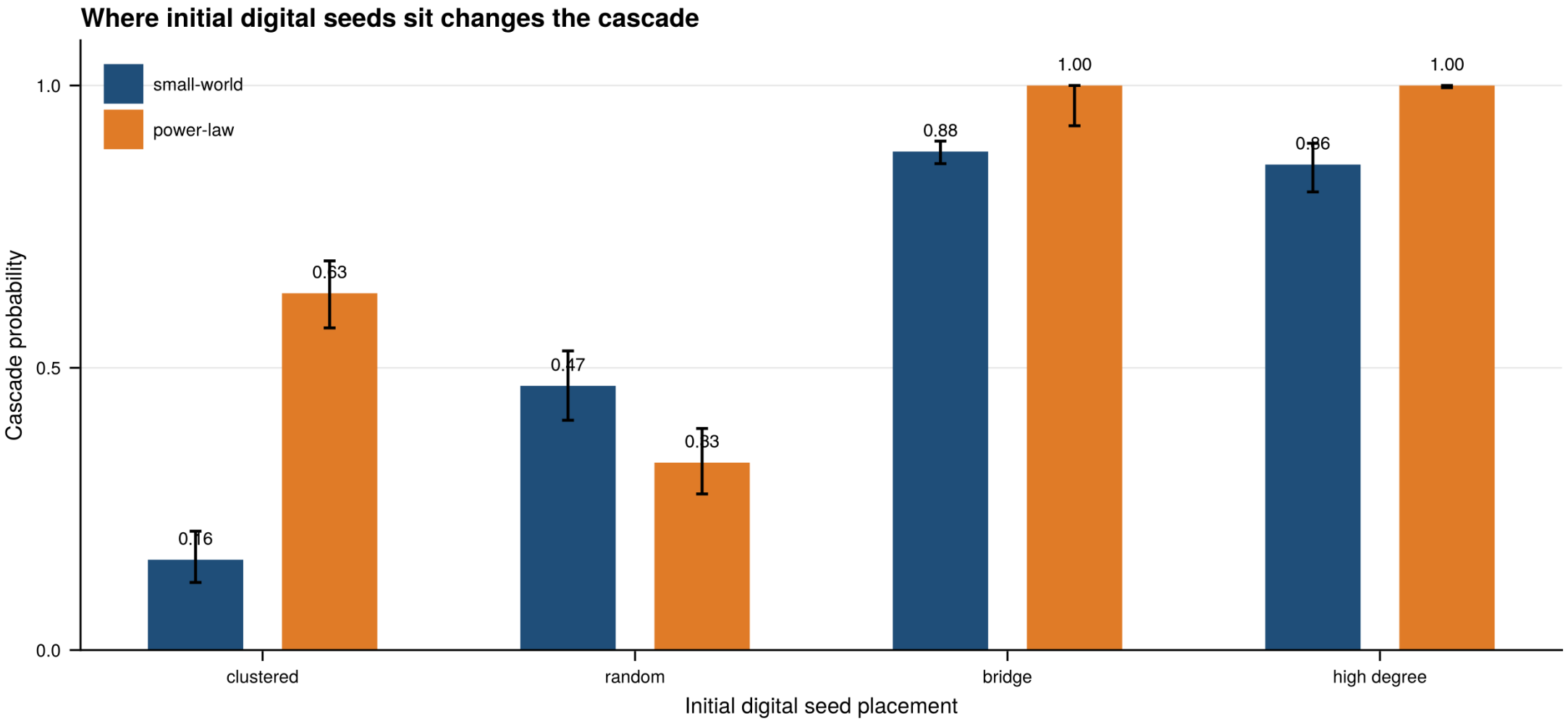

**Figure 6.** *Experiment D: cascade probability by initial-seed placement and network family (N = 1,000; Wilson 95% CIs).*

**Table 5.** *Experiment D — initial digital seed placement (cascade probability).*

| Family | clustered | random | bridge | hub |
|---|---|---|---|---|
| small-world (β = 0.4) | 0.16 | 0.47 | 0.88 | 0.86 |
| power-law (m = 2) | 0.63 | 0.33 | 1.00 | 1.00 |

**Table 6.** *Experiment E — 2×2 of initial-seed × pioneer placement (cascade probability; median $T_{50}$; n = 1000 per cell).*

| Seed placement | Pioneer placement | Cascade | $T_{50}$ |
|---|---|---|---|
| random | random | 0.51 | 54 |
| random | hub/bridge | 1.00 | 24 |
| hub/bridge | random | 0.92 | 35 |
| hub/bridge | hub/bridge | 1.00 | 20 |

## 6.5 Robustness of the mechanism: adoption rule and replacement entrants

Two robustness checks probe whether the headline mechanism is an artefact of the exposure definition or of the replacement-entry rule.

**How much reinforcement does the rule require? (Experiment G.)** We re-ran the focal Experiment A cell and the β = 0.4 transition cell (R = 250 each) under stricter, reinforcement-requiring adoption rules (Figure 7, Table 7). In this deliberately sparse calibration (mean degree 4) the result is sharply bounded: the original performance-conditioned rule cascades (0.99 at Exp A; 0.46 at the transition) and an absolute single-neighbour rule also cascades (1.00), but every variant that requires two or more successful neighbours, or an all-neighbour denominator, collapses to 0.00. At mean degree 4 the better-performing set is so small that the fractional rule reduces to a single-exemplar trigger, and a strict reinforcement gate has no chance to ignite. This is a property of the sparse calibration, not of the rule: real firm observation/peer groups are not of size four. Section 6.6 shows that once the neighbourhood is realistic and human-scale, the very same rule becomes genuine reinforcement-based complex contagion — adoptions are triggered by many successful exemplars and the strict ≥2 rule survives. The contrast with the hazard and frequency-based nulls holds throughout (neither cascades under any specification).

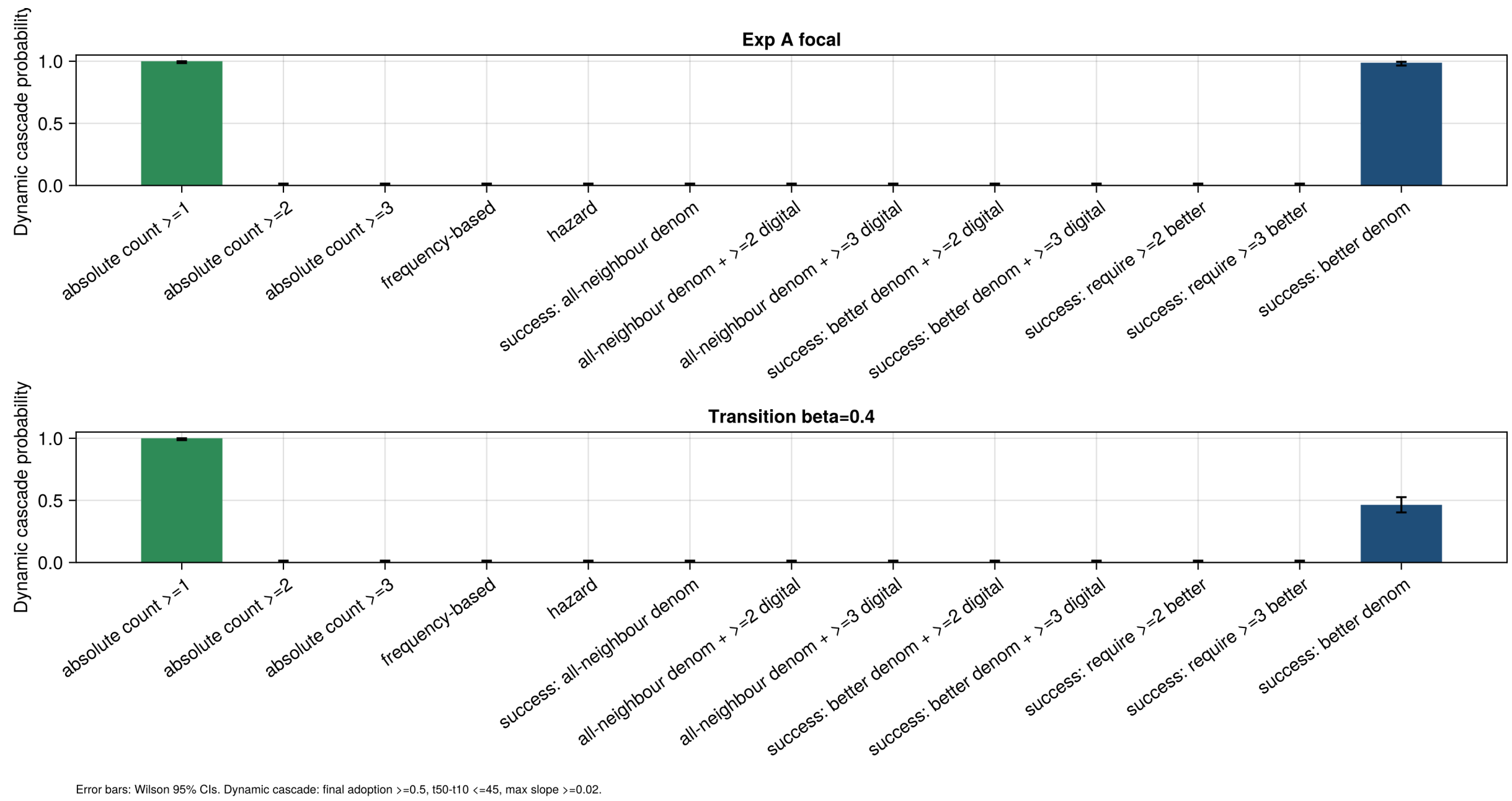

**Figure 7.** *Experiment G — success-rule robustness (R = 250). Cascade probability under the original performance-conditioned rule, single-neighbour count, and stricter reinforcement variants, in the Exp A focal and β = 0.4 transition settings. Only the single-exemplar rules cascade; all minimum-count and all-neighbour-denominator variants are at zero. Wilson 95% CIs.*

**Table 7.** *Experiment G — adoption-rule robustness (cascade probability, Wilson 95% CI; R = 250).*

| Adoption rule variant | Exp A focal | β = 0.4 transition |
|---|---|---|
| Original (digital / better-performing) | 0.99 [0.97, 1.00] | 0.46 [0.40, 0.53] |
| Count ≥ 1 better-performing digital | 1.00 [0.98, 1.00] | 1.00 [0.98, 1.00] |
| Require ≥ 2 successful digital | 0.00 [0.00, 0.02] | 0.00 [0.00, 0.02] |
| Require ≥ 3 successful digital | 0.00 [0.00, 0.02] | 0.00 [0.00, 0.02] |
| All-neighbour denominator | 0.00 [0.00, 0.02] | 0.00 [0.00, 0.02] |
| Require ≥ 2 better-performing (any) | 0.00 [0.00, 0.02] | 0.00 [0.00, 0.02] |
| Hazard / frequency-based nulls | 0.00 / 0.00 | 0.00 / 0.00 |

**Does the replacement rule matter? (Experiment I.)** We ablated the two replacement-entry rules (Section 3.9) across mechanism, transition, and placement cells (R = 250; Figure 8, Table 8). The cascade probability barely moves between traditional-only and Bernoulli-seed entrants: 1.00 vs 0.99 for the Experiment A performance-conditioned cell, 0.00 vs 0.00 for hazard, 0.56 vs 0.58 for the β = 0.4 mixture (π = 5%), 0.44 vs 0.52 for random seed/random pioneer placement, and 1.00 vs 1.00 for hub/bridge seed and pioneer placement. The continuing-reseeding behaviour in the production code is therefore not driving the results; the manuscript baseline is the traditional-only rule, and the conclusions are robust to the alternative.

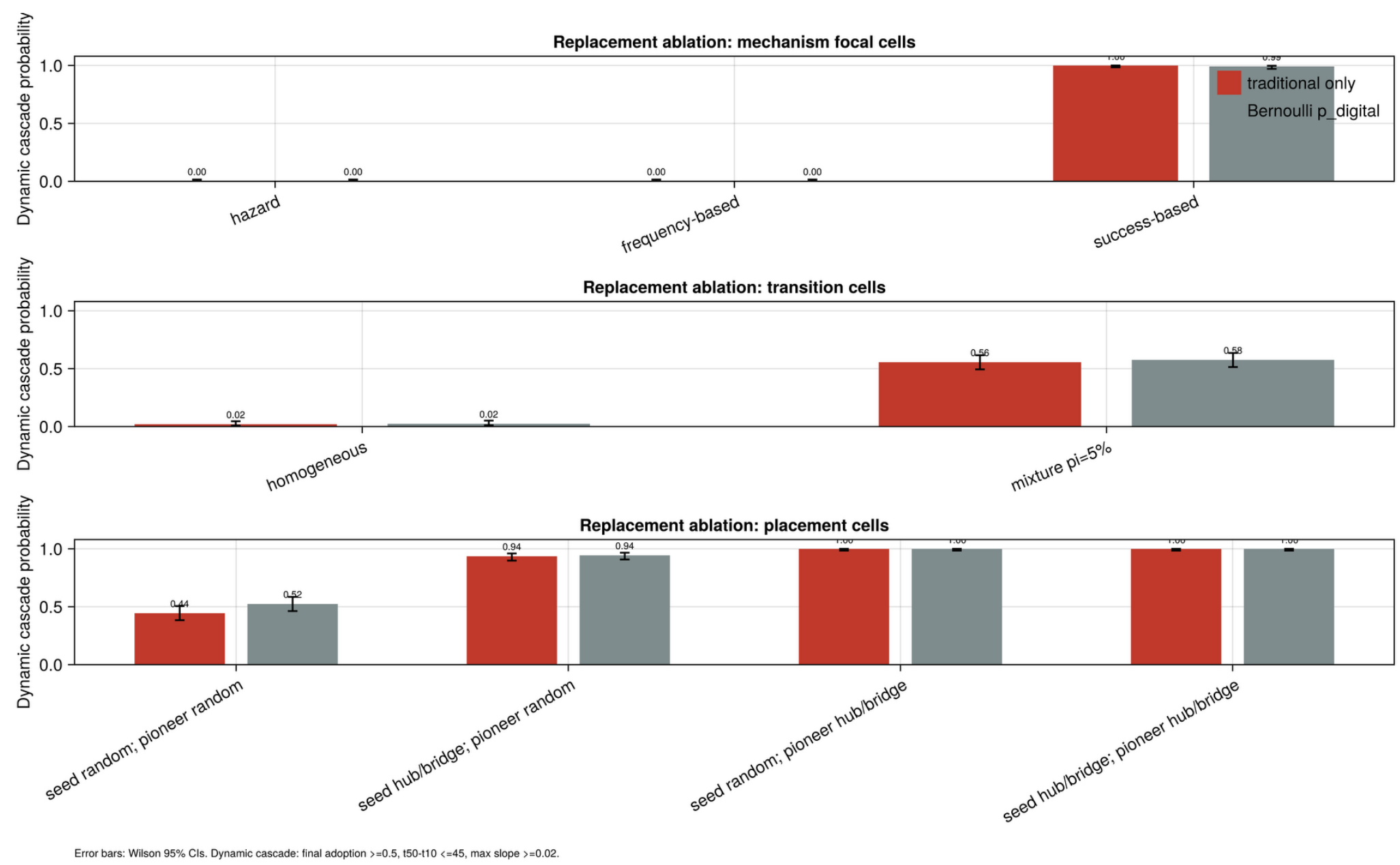

**Figure 8.** *Experiment I — replacement-entry ablation (R = 250). Cascade probability under traditional-only versus Bernoulli-seed replacement entrants across mechanism, transition, and placement cells. Wilson 95% CIs.*

**Table 8.** *Experiment I — replacement-entry ablation (cascade probability; R = 250).*

| Cell | Traditional-only | Bernoulli-seed |
|---|---|---|
| Exp A — performance-conditioned | 1.00 | 0.99 |
| Exp A — hazard | 0.00 | 0.00 |
| Transition β = 0.4 — mixture π = 5% | 0.56 | 0.58 |
| Random seed / random pioneer | 0.44 | 0.52 |
| Hub-bridge seed / hub-bridge pioneer | 1.00 | 1.00 |

**Parameter robustness (Experiment F).** Sweeping the model around the focal scenario (Appendix E), the cascade vanishes at a 1.5–2× digital premium, emerges at 3× when digital variance is low (0.60), and is present at 5× and 10× (0.89 and 0.85 at low variance): it requires a sizeable but not extreme premium. Cascade probability rises monotonically with seed size (0.06 → 0.49 → 0.92) and is preserved across return distributions (normal, truncated normal, lognormal: 0.42–0.80), benchmark definitions, synchronous versus asynchronous updates (0.46 vs 0.78), and absorbing versus non-absorbing adoption (0.47 vs 0.55). Notably, lower digital variance strengthens the cascade, so high variance is best read as noisy local inference rather than as a cascade amplifier.

## 6.6 Realistic, human-scale degree: genuine complex contagion, ignition, and bounded scale

The sparse single-exemplar regime of Section 6.5 is not realistic: firms observe peer groups of tens to a few hundred, not four. Experiments K and L re-run the model at realistic, degree-controlled neighbourhood sizes (k = 10–150), recalibrating the phase boundary at each degree, and add a direct diagnostic — the number of successful digital exemplars a firm actually observes at the moment it adopts. The picture changes decisively and the complex-contagion reading is recovered, now demonstrated rather than assumed.

**Adoption is reinforcement-driven.** At realistic degree, firms adopt only after observing many successful digital exemplars: the median number of successful (better-performing, digital) neighbours at adoption is about 9 at k = 20 and 25 at k = 50 (Figure 11), with the exposure share at adoption ≈ 1.0 — a firm switches only when essentially all of its better-performing peers are digital. This is the defining signature of complex contagion: adoption requires broad local reinforcement, not a single exposure.

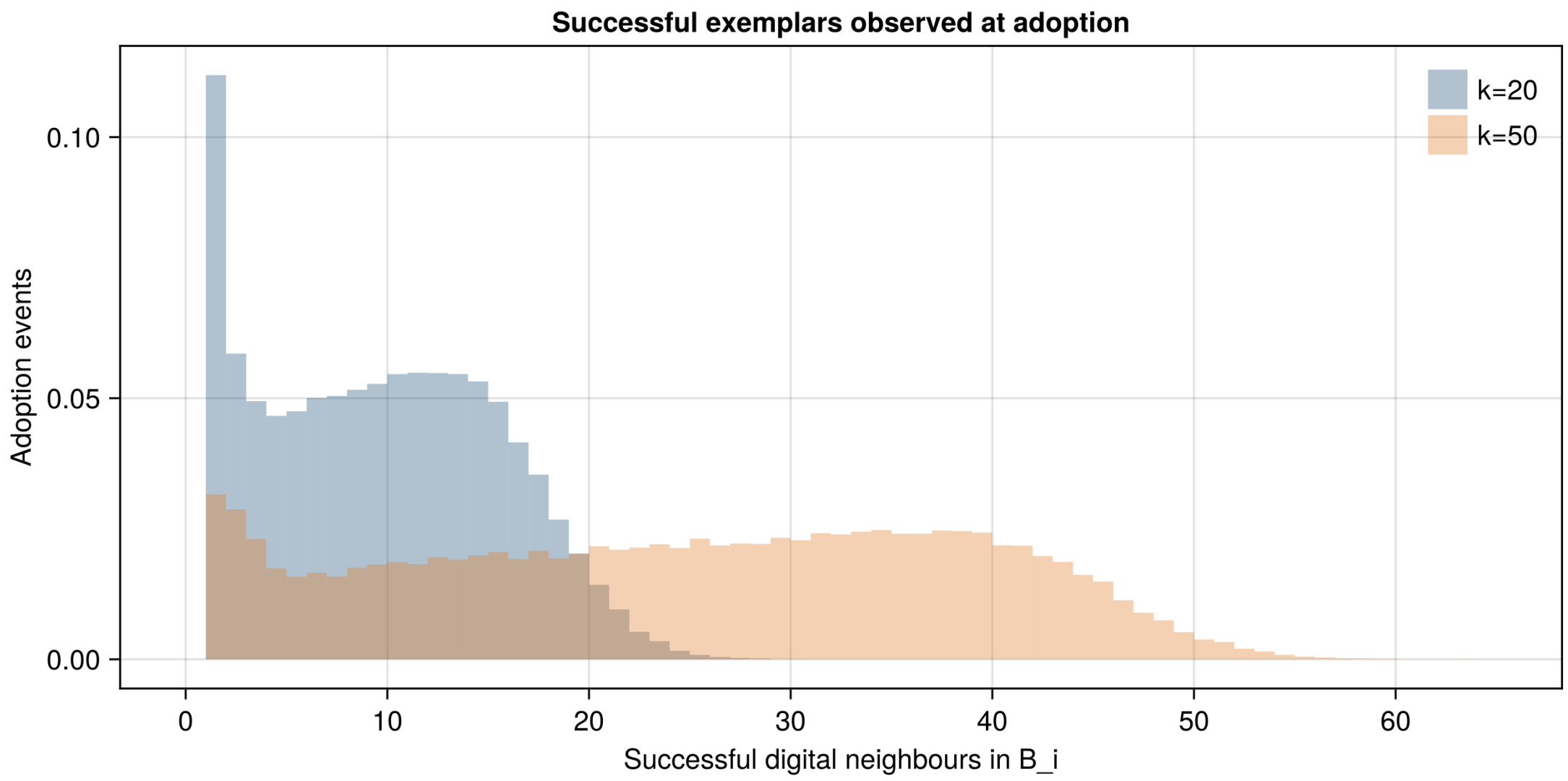

**Figure 11.** *Experiment L — number of successful digital exemplars observed at the moment of adoption, at realistic degree. Medians ≈ 9 (k = 20) and ≈ 25 (k = 50); adoption requires reinforcement from many exemplars, not one.*

**Reinforcement survives a strict minimum-count rule, once ignition is adequate.** The earlier collapse of the ≥2 rule (Section 6.5) was an ignition artefact of a single-firm seed: a strict reinforcement gate cannot bootstrap when no firm begins with two digital neighbours. With an adequate — and preferentially clustered — seed, the explicit ≥2 (and even ≥3) successful-digital rule ignites and cascades at realistic degree (Figure 12). For example, at k = 50 the ≥2 rule reaches cascade probability ≈ 0.53–0.68 and the ≥3 rule ≈ 0.37–0.43 once the seed exceeds roughly 0.5–2% of firms, whereas the original fractional rule ignites from a 0.1% seed. That reinforcement-based adoption needs a critical seed mass to ignite — and benefits from clustered or bridge placement — is exactly the Centola–Watts signature of complex contagion, not a weakness of it.

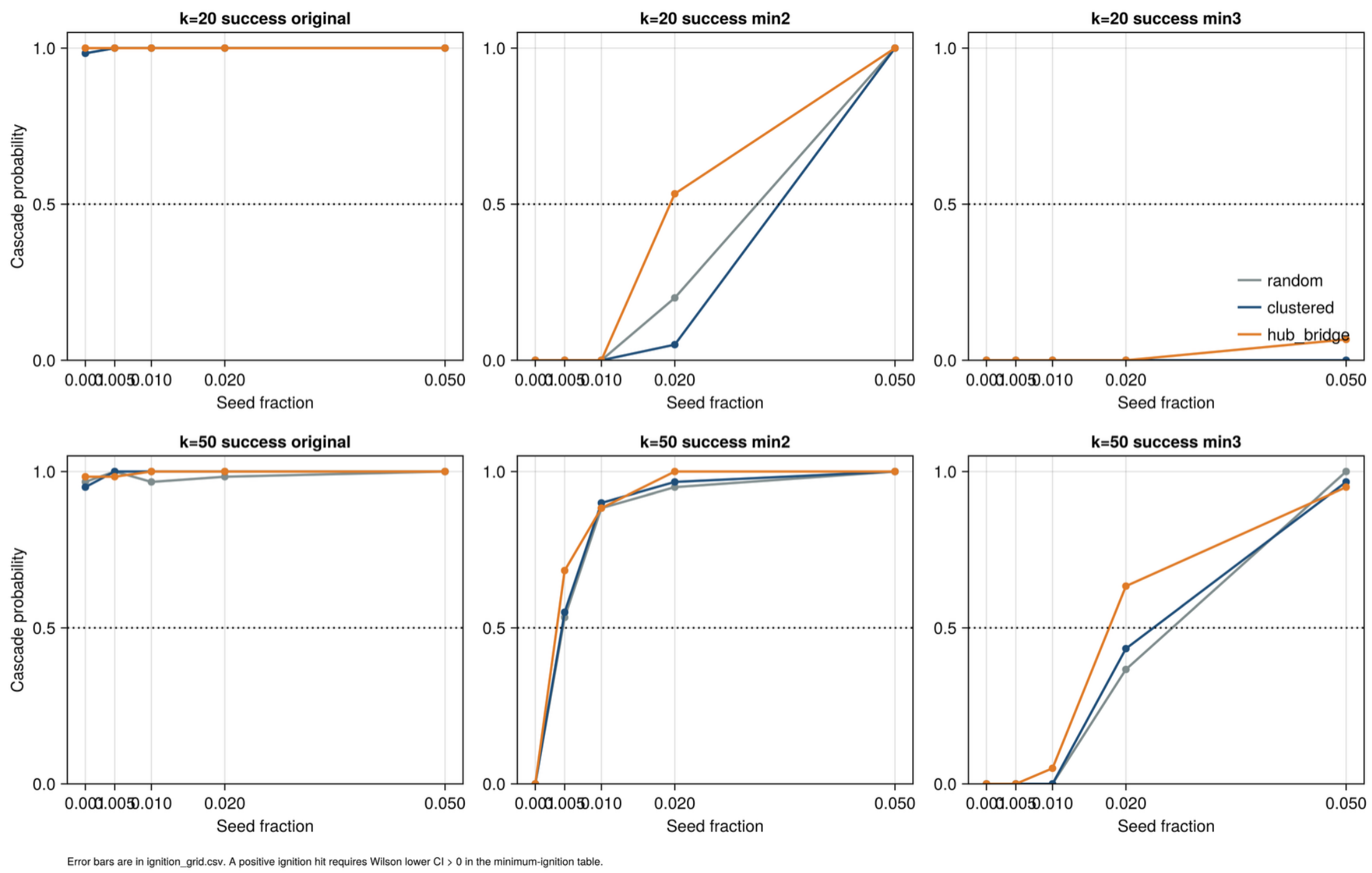

**Figure 12.** *Experiment L — ignition grid: cascade probability for the ≥1, ≥2, and ≥3 successful-digital rules across seed size and placement at realistic degree. The strict reinforcement rules ignite once the seed is large enough and clustered/bridge-placed. Wilson 95% CIs.*

**The mechanism is bounded to human scale.** Cascade strength is non-monotonic in neighbourhood size (Figure 13). It is robust across human-scale degrees — cascade probability is 1.0 at k = 10, 15, 30, and 50 — and then collapses to 0.0 by k = 100–150 and beyond. This is consistent with bounded, fast-and-frugal cognition (Gigerenzer) and with the Dunbar limit on workable group size: a firm cannot meaningfully observe and socially validate against an unbounded peer set, and very large clusters subdivide into human-manageable units. A mechanism that operates within human-scale neighbourhoods and weakens past them is therefore expected on the theory, not a defect; the model places the operative ceiling at roughly k ≈ 50–100, in the neighbourhood of human-scale bounds.

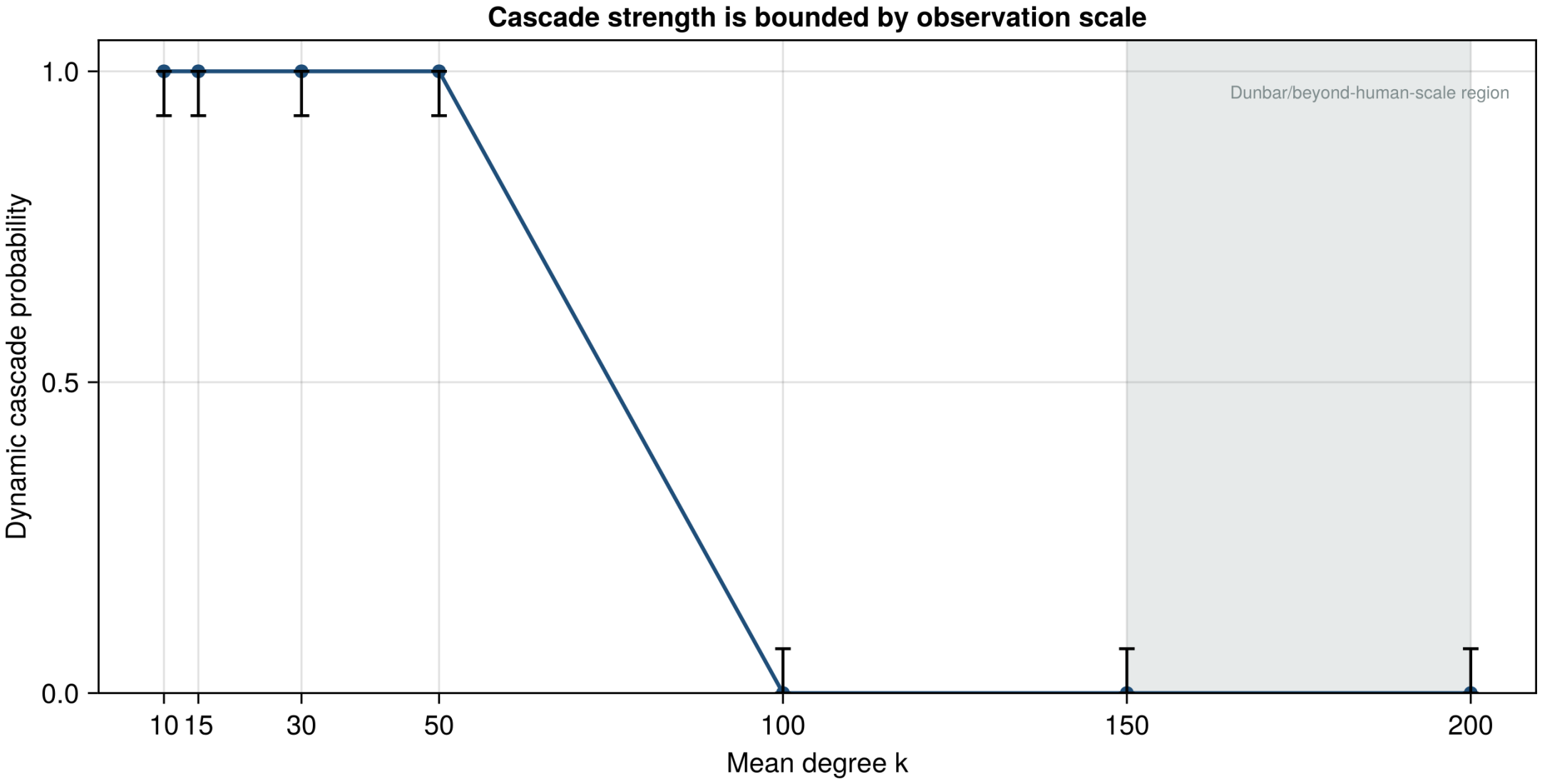

**Figure 13.** *Experiment L — cascade probability versus observation degree (Dunbar region marked). The cascade is robust at human-scale neighbourhoods (k ≤ 50) and collapses at large neighbourhoods (k ≥ 100), consistent with bounded cognition.*

Taken together, Experiments K and L recover the complex-contagion reading on firmer ground than the original model: at realistic, human-scale degree the performance-conditioned rule is genuine reinforcement-based complex contagion (many exemplars per adoption), it requires a critical, well-placed seed to ignite, and it operates within the bounded neighbourhoods that human cognition and organizational structure actually sustain.

## 6.7 Sensitivity to the cascade definition

The dynamic cascade definition (final adoption ≥ 0.5, transition ≤ W = 45, slope ≥ s_min = 0.02) is conventional, so Experiment H recomputes cascade classifications from the existing trajectories over a grid of final cutoff {0.5, 0.7}, W {30, 45, 60}, and s_min {0.01, 0.015, 0.02, 0.025, 0.03} (Figure 9, Appendix H). The qualitative conclusions are stable. The hazard and frequency-based rules are never reclassified as cascades, at any point in the grid (cascade probability remains 0.00). Performance-conditioned imitation remains clearly distinguished from slow accumulation: its Experiment A cascade probability is 0.99 at the baseline, 1.00 under the loosest setting (W = 60, s_min = 0.01), and 0.60 under the strictest (final ≥ 0.7, W = 30, s_min = 0.03). The absolute probabilities of intermediate transition cells shift with W and s_min, as expected, but no headline ordering reverses. We retain W = 45 and s_min = 0.02 as a middle-of-grid baseline and report the full grid in the appendix.

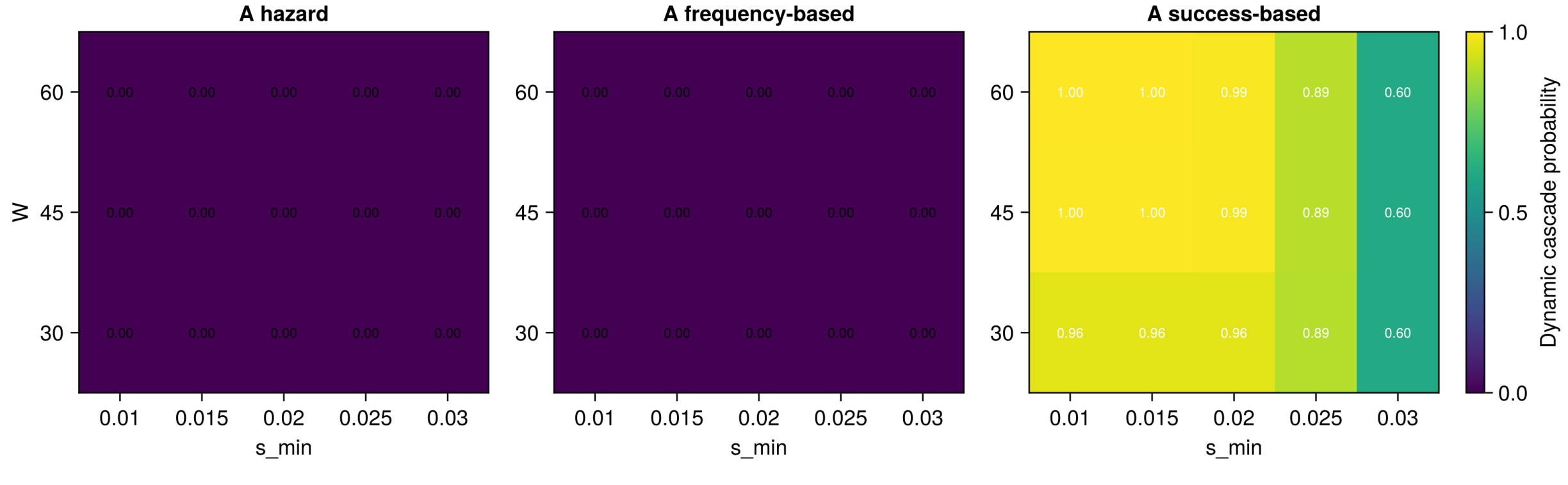

**Figure 9.** *Experiment H — cascade-metric sensitivity. Cascade probability for the Experiment A rules (and selected cells) recomputed across the W × s_min × final-cutoff grid. Hazard and frequency-based rules remain at zero throughout.*

## 6.8 Parameter-space scope

Because the baseline regime is deliberately placed near the cascade transition, Experiment J maps cascade probability over three parameter planes (R = 100 per cell; production replacement rule) to show where the mechanism operates and where it fails (Figure 10). Over β × misfit share, cascades are absent at low connectivity (β = 0.2, π = 0.05 gives 0.02; π = 0.10 gives 0.07) and cross a frontier as connectivity rises (β = 0.4/0.7/1.0 at π = 0.05 give 0.49/0.83/0.91). Over $\bar{\theta}$ × seed fraction at β = 0.4, the seed fraction is pivotal ($\bar{\theta}$ = 0.70 gives 0.15, 0.43, 0.93 at seed fractions 0.003, 0.005, 0.010). Over digital premium × variance, the cascade needs a sizeable premium and is stronger at lower variance (premium 10× gives 0.51 at high variance and 0.93 at low variance; the stricter ≥2-neighbour rule gives 0.00 even at 10×, consistent with Experiment G). The mechanism thus holds over a bounded region: it requires sufficient network reach, a sizeable digital premium, and enough well-placed seeds or low-threshold pioneers, and it fails outside that region.

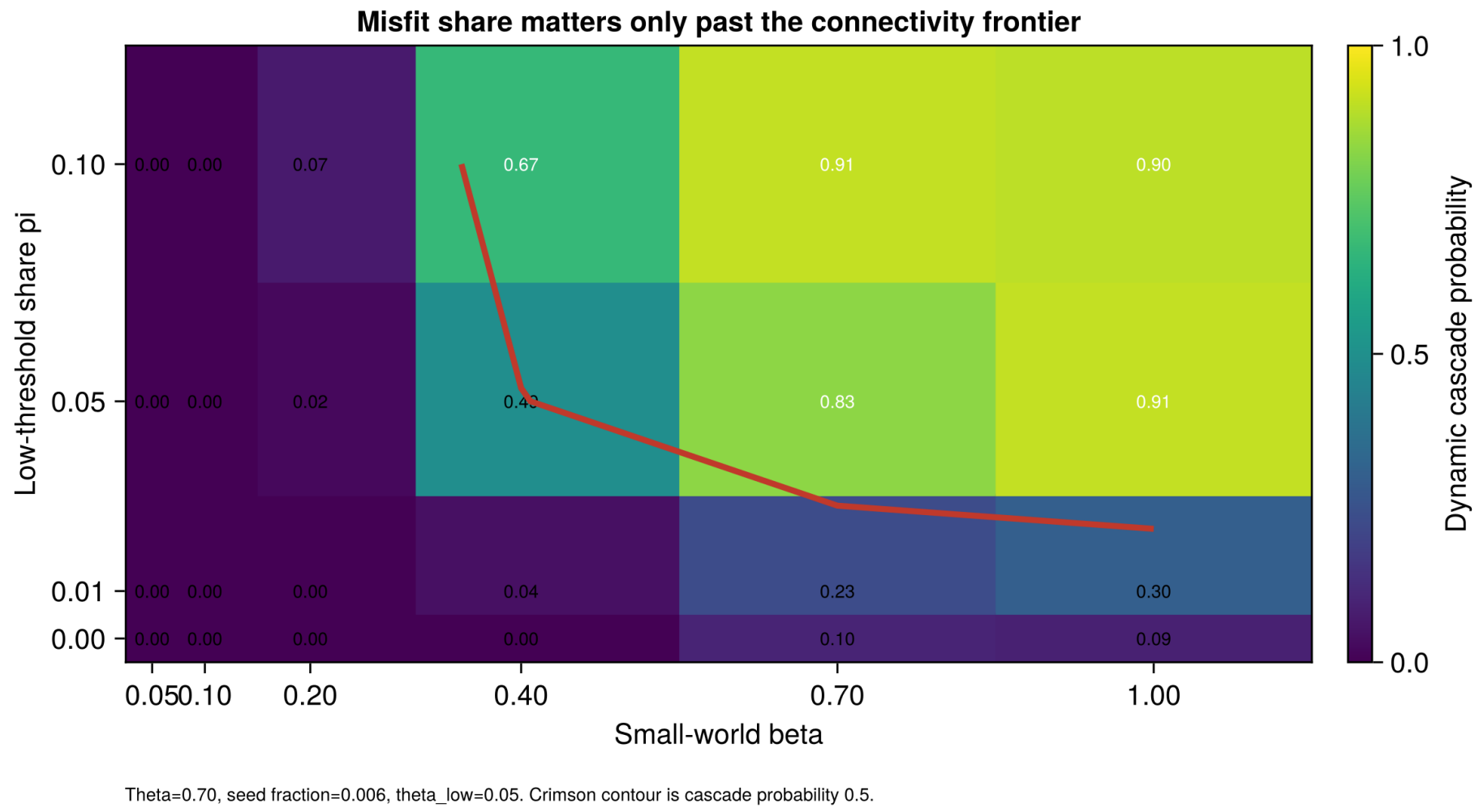

**Figure 10.** *Experiment J — parameter-space map of cascade probability over small-world β and misfit share π (R = 100; production replacement rule). Companion $\bar{\vartheta}$ × seed-fraction and premium × variance maps are in Appendix (parameter-space).*

## 7. Verification, validation, and reproducibility

**Code verification.** The implementation includes a unit-test suite covering the adoption rules, the network generators (including the degree-conserving rewiring), threshold draws, and the cascade metric; an explicit audit of the implementation (Appendix J) documents each submodel and surfaced the replacement-entry discrepancy resolved in Section 3.9.

**Model documentation.** The model is documented with a full ODD protocol (Appendix A) and pseudocode (Appendix, Reproducibility), sufficient for reimplementation.

**Stylized-facts validation.** As corroboration of intended behaviour rather than empirical prediction, the model reproduces four stylized facts of technology diffusion: slow adoption despite high expected returns, sudden endogenous takeoff, clustered adoption, and strong sensitivity to visible successful exemplars (Appendix F).

**Sensitivity and robustness.** Experiments F–L provide parameter sensitivity, adoption-rule robustness, cascade-metric sensitivity, replacement-entry ablation, parameter-space scoping, and the realistic-degree / ignition / bounded-scale analyses, with Wilson confidence intervals throughout.

**Reproducibility package.** Every experiment writes per-run, per-period, and summary CSVs and a metadata JSON recording the master seed, per-run seeds, settings, and package versions; figures are regenerated from the CSVs. Software versions, seeds, run commands, and expected outputs are listed in the reproducibility appendix. The complete code, ODD, and an archived release (with DOI) are deposited in a public repository; the link is withheld here for anonymous review and has been provided to the editor.

**Limits of validation.** Validation here is fitness for the explanatory purpose of isolating a mechanism. The model is not calibrated to any firm population and makes no industry-level predictions; its outputs are qualitative regularities and conditional statements, not forecasts.

## 8. Discussion

The results cohere into one scoped argument about the calibrated regime studied here. In this model, dynamic cascades are produced by performance-conditioned imitation but not by an exogenous hazard or by frequency-based, bandwagon imitation (Experiment A): firms act on the visible success of digital peers, not on prevalence. The spread is gated by network reach (Experiment B): it requires short average path length and low clustering, supplied by small-world rewiring, scale-free hubs, or random short paths, with a sharp degree-controlled tipping point at $\beta^* \approx 0.41$; because these structural properties are collinear, the topology result is best interpreted as a network-reach condition rather than an isolated path-length effect. Given enough reach, the shape of the receptiveness distribution matters (Experiment C): concentrating receptiveness in the lower tail can trigger cascades where homogeneity cannot, but equal-effort matching shows this is a statement about the value of lower-tail heterogeneity rather than a blanket superiority over lowering the mean. Placement strongly conditions the outcome (Experiments D–E): both initial adopters and low-threshold pioneers are far more effective on hubs and bridges than scattered or clustered.

The robustness programme sharpens rather than undercuts the claim. At a deliberately sparse mean degree of four the rule reduces to a single-exemplar trigger and a strict reinforcement requirement cannot ignite (Experiment G) — but that is a property of an unrealistic neighbourhood size. At realistic, human-scale degree (Experiments K and L) the same rule is genuine reinforcement-based complex contagion: adoptions are triggered by many successful exemplars (median ≈ 9–25), the explicit ≥2 rule survives once ignition is given an adequate, clustered seed (the Centola–Watts critical-mass signature), and the cascade operates within human-scale neighbourhoods and weakens beyond them, consistent with bounded cognition (Gigerenzer) and the Dunbar limit. The mechanism is also robust to the replacement-entry rule (Experiment I) and to the cascade-metric choices (Experiment H), and it operates over a bounded but non-trivial region of parameter space (Experiment J). The fragility at degree four is thus informative: complex contagion among firms is real but requires the kind of bounded, reinforcing observation neighbourhood that organizations actually inhabit.

The theoretical contribution is the firm-level combination, not the ingredients. Thresholds, social learning, and cascades are well established; what this model isolates is how a performance-conditioned imitation rule interacts with threshold heterogeneity, network reach, and seed/pioneer placement to separate mechanisms that discussions of digital transformation usually conflate. The policy reading is correspondingly modest: the simulations suggest that cultivating a small number of receptive firms and connecting them across community boundaries, so their success becomes locally visible, may have more leverage than a broad shift in average receptiveness — but only where the network supplies reach and the technology's advantage is large. The model indicates where interventions may have leverage; it does not forecast adoption in any particular industry.

## 9. Limitations

Several limitations bound the interpretation of these results.

- Abstract network structure. The observation networks are stylized (small-world, scale-free, and random families); they are not derived from real inter-firm networks, and the inter-community structure is implicit rather than an explicit, tunable block model.
- Large digital premium. The baseline digital premium is tenfold by construction. Experiments F and J show the cascade requires a sizeable premium (emerging around 3–5×) and is absent at small premia, so the premium should be read as a strong signal condition rather than an empirical estimate.
- Phase-boundary calibration. The baseline regime is deliberately placed near the cascade transition so treatment contrasts are visible. This is a design choice, not a claim of typicality; Experiment J maps where the mechanism holds and where it fails.

- Degree-dependent reinforcement. At a sparse mean degree of four the rule reduces to a single-exemplar trigger and a strict ≥2 requirement cannot ignite (Experiment G); the genuine reinforcement-based complex-contagion reading holds at realistic, human-scale degree (Experiments K and L), where adoptions involve many exemplars and the ≥2 rule survives under adequate ignition. The complex-contagion claim is therefore explicitly scoped to realistic neighbourhood sizes.
- Ignition requires a critical seed. Reinforcement-based adoption needs an adequate, preferentially clustered initial seed to ignite; from a single seed a strict reinforcement rule cannot start. This is theory-consistent (Centola–Watts) but means the cascade is conditional on seed mass and placement, not automatic.
- Bounded observation scale. The cascade operates within human-scale neighbourhoods and weakens past roughly $k \approx 50$–100; the precise ceiling is a model output, interpreted in the spirit of bounded cognition and the Dunbar limit rather than as a calibrated value.
- Threshold interpretation. The threshold is organizational receptiveness, of which culture is one possible component among risk tolerance, absorptive capacity, capital constraints, and managerial attention; we do not claim a specifically cultural reading.
- Topology attribution. Average path length, clustering, and degree are collinear across our network families, so the cross-family regression cannot isolate a single statistic; the causal topology claim rests on the degree-controlled sweep and is framed as network reach.
- Absorbing adoption and simplified social comparison. Adoption is absorbing in the baseline (a non-absorbing variant is tested in robustness), and social comparison is a fast-and-frugal heuristic rather than an inferential learning model.
- No empirical calibration. The model is explanatory; it is not fitted to firm-level data and makes no industry-level forecasts.

## 10. Conclusion

In an agent-based model of firms that adopt a digital technology by imitating successful neighbours, system-wide adoption is governed by the interaction of the imitation logic, the reach of the observation network, the shape of the receptiveness distribution, and the placement of seeds and pioneers. Dynamic cascades are produced by performance-conditioned imitation but not by an exogenous hazard or by frequency-based imitation; they require sufficient network reach; and, given reach, lower-tail threshold heterogeneity — especially a few low-threshold pioneers on hubs and bridges — can tip the system without any change in the average firm. At realistic, human-scale observation neighbourhoods the mechanism is genuine reinforcement-based complex contagion: adoption is driven by many successful exemplars, it needs a critical, well-placed seed to ignite, and — consistent with bounded cognition and the Dunbar limit — it operates within human-scale groups and fades beyond them. The robustness programme makes these conditions precise rather than weakening them, and the result is robust to the replacement-entry rule, the cascade definition, and a bounded but non-trivial region of parameter space.

The practical lesson is the one the model was built to test, now on firmer ground: broad cultural change is not a prerequisite for technological disruption. A small minority of receptive firms — a few misfits — placed where the network can carry their success, can be enough to tip the system. In that precise and bounded sense, a few misfits can change the world.

## Appendix A: Full ODD protocol

### A.1 Overview

Purpose (Section 3.1): explain when firm-level digital adoption becomes a cascade and whether lower-tail receptiveness can substitute for a mean shift. Entities: firms on a fixed undirected network. State variables are listed in Table A1. Scales: discrete time, N = 1,000 firms, T = 100 periods.

**Table A1.** *State variables.*

| Variable | Meaning |
|---|---|
| $d_i(t) \in \{0,1\}$ | Digital status; absorbing once 1 (non-absorbing variant in robustness). |
| $v_i(t) > 0$ | Valuation, multiplicative stochastic returns. |
| $\theta_i \in [0,1]$ | Adoption threshold = organizational receptiveness. |
| $pioneer_i \in \{0,1\}$ | Low-threshold pioneer flag (misfit); independent of initial digital status. |
| $a_i(t)$ | Age since entry. |
| node | Fixed network position determining the neighbour set $N_i$. |

### A.2 Design concepts

Basic principle: local social validation via a fast-and-frugal imitation heuristic. Emergence: cascades and their timing are not imposed. Adaptation: status switching. Objectives/learning: observational, not inferential. Interaction: network-mediated. Stochasticity: network realisation, seed and pioneer placement, return shocks, threshold draws. Observation: adoption share, $T_{10}/T_{50}/T_{90}$, dynamic cascade probability with Wilson intervals, transition time, and maximum slope.

### A.3 Details

Initialization: common initial valuation, age zero; a 0.3% initial digital seed; thresholds drawn per treatment (Section 3.6). Input data: none. Submodels: valuation dynamics (Eq. 1–2), the two imitation rules and hazard null (Eq. 3–5), the receptiveness distribution (Eq. 6), and the degree-controlled network generators (Section 3.7). Scheduling is synchronous by default; an asynchronous variant is tested in robustness.

## Appendix B: Calibration

The cascade regime was located by sweeping the mean threshold $\bar{\theta}$ from 0.20 to 0.70 on a small-world $\beta = 0.2$ network and comparing the homogeneous baseline with a mean-matched mixture ($\pi = 0.05$). With seed fraction 0.005 and degree 6, even $\bar{\theta} = 0.70$ cascaded too reliably; hardening to seed fraction 0.003 and degree 4 placed the homogeneous baseline at the phase boundary (homogeneous cascade probability ≈ 0.30 at $\bar{\theta} = 0.70$, mixture ≈ 0.80). The reported experiments use this calibrated regime so that treatment effects are visible rather than saturated.

## Appendix C: Degree-controlled topology and regression

Small-world networks use degree-conserving Watts–Strogatz rewiring; realised mean degree is 4.0 for every $\beta$. Structural metrics per family are recorded in the run CSVs. The bootstrapped small-world tipping point is $\beta^* = 0.410$ with a tight 95% CI [0.392, 0.428]. A logistic regression of the cascade indicator on structural covariates pooled across families yields the coefficients in Table C1. The covariates are strongly collinear across our network families — as $\beta$ rises, average path length and clustering fall together while the families differ systematically in degree — so the regression cannot uniquely attribute the effect to a single statistic. In the full sample the signed contributions load on clustering (strongly negative) and mean degree (positive), with average path length not separately identified (odds ratio ≈ 1.0). We therefore rely on the degree-controlled small-world sweep (Figure 2a) for the causal topology claim — there, with degree fixed, cascade probability jumps from near zero to near one over a narrow $\beta$ range — and we describe the controlling factor as network reach (short paths, low clustering, hubs) rather than any single metric.

**Table C1.** *Logistic regression of cascade on network structure (pooled across families; full mode). Coefficients reflect collinear covariates and should not be read as isolated marginal effects.*

| Term | Estimate | Odds ratio |
|---|---|---|
| Intercept | −23.01 | — |
| Average path length | 0.003 | 1.003 (n.s.) |
| Mean degree | 6.55 | 698 |
| Clustering | −22.35 | $\approx 2\times10^{-10}$ |
| Degree variance | −0.07 | 0.93 |
| Family = power-law | −0.72 | 0.49 |
| Family = small-world | −0.95 | 0.39 |

## Appendix D: Equal-effort culture comparison

To test whether the lower-tail advantage is an artifact of unequal intervention size, treatments were matched on a common budget: equal L1 distance of the threshold distribution from the homogeneous baseline, and equal expected mass below a critical threshold. Cascade probabilities are in Table D1 and Figure D1. At the smallest L1 budget the mixture leads narrowly; at the larger L1 budget and under both mass-below-threshold budgets, a matched lowering of the mean (and raising variance) is as effective or more. The defensible conclusion is that lower-tail heterogeneity is a powerful and often sufficient lever, but not uniformly superior to an equal-effort mean shift; the ordering depends on the effort metric.

**Table D1.** *Equal-effort comparison (cascade probability; N = 1,000; n = 1,000–1,050 per cell).*

| Budget | Lower mean | Higher variance | Mixture |
|---|---|---|---|
| L1 distance = 0.02 | 0.37 | 0.34 | 0.40 |
| L1 distance = 0.05 | 1.00 | 0.63 | 0.74 |
| Mass below θ_crit = 0.03 | 1.00 | 1.00 | 0.57 |
| Mass below θ_crit = 0.05 | 1.00 | 1.00 | 0.87 |

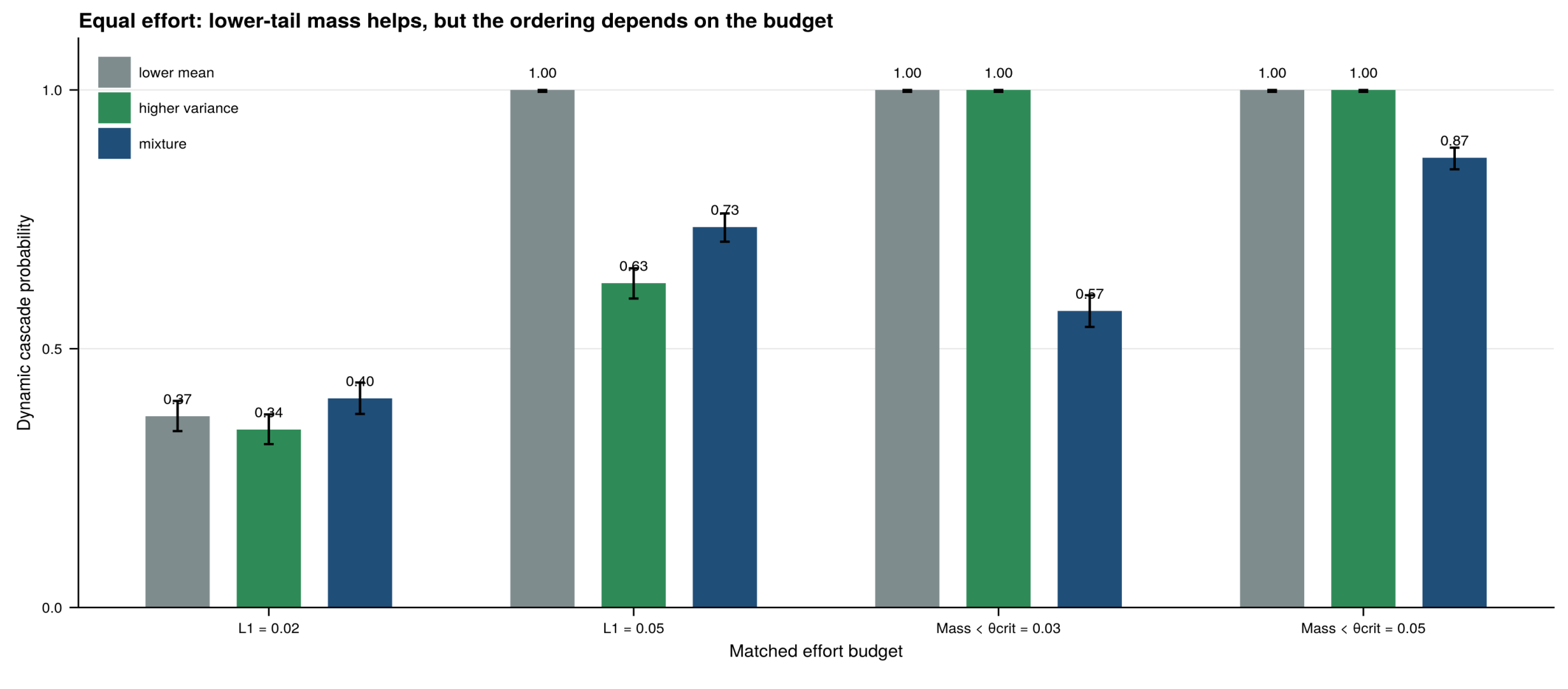

**Figure D1.** *Equal-effort threshold shifts (N = 1,000; Wilson 95% CIs). Lower-tail mass helps, but the ordering of treatments depends on the effort budget.*

## Appendix E: Robustness

Table E1 summarises the robustness sweeps. The cascade requires a sizeable digital premium (absent at 1.5–2×, emerging at 3×, firm at 5–10×), strengthens monotonically with seed size, and is preserved across return distributions, benchmark definitions, update schedules, and non-absorbing adoption.

**Table E1.** *Robustness sweeps (cascade probability; N = 1,000; low-variance digital returns unless noted).*

| Axis | Condition | Cascade prob. |
|---|---|---|
| Digital premium | 1.5× / 2× / 3× / 5× / 10× | 0.00 / 0.08 / 0.60 / 0.89 / 0.85 |
| Seed fraction | low → high | 0.06 / 0.49 / 0.92 |
| Return distribution | normal / truncated / lognormal | 0.42 / 0.78 / 0.80 |
| Update schedule | synchronous / asynchronous | 0.46 / 0.78 |
| Adoption | absorbing / non-absorbing | 0.47 / 0.55 |

## Appendix F: Stylized facts

As corroboration of intended behaviour (not empirical prediction), the model reproduces four stylized facts of technology diffusion (Table F1): slow adoption despite high expected returns; sudden endogenous takeoff; clustered adoption; and strong sensitivity to visible successful exemplars.

**Table F1.** *Stylized-facts check (N = 1,000).*

| Stylized fact | Median final A | Cascade prob. |
|---|---|---|
| Slow adoption despite high returns | 0.07 | 0.00 |
| Sudden endogenous takeoff | 0.30 | 0.00* |
| Clustered adoption | 0.04 | 0.00 |
| Sensitivity to visible exemplars | 0.80 | 0.80 |

***** The takeoff scenario produces a late but real acceleration; under the strict dynamic window it is borderline, illustrating the conservatism of the cascade metric.

## Appendix G: Success-rule robustness (Experiment G)

Experiment G tests whether the performance-conditioned result depends on the exposure definition, by re-running the focal Experiment A cell and the $\beta = 0.4$ transition cell ($R = 250$) under a family of stricter and alternative adoption rules. Full per-variant cascade probabilities with Wilson 95% confidence intervals are in Table 7 (Section 6.5) and in the replication CSVs. The key finding is that, at this unrealistic mean degree of four, the cascade survives only when a single successful exemplar can trigger adoption (the original better-performing-denominator rule and the absolute count-≥1 rule); it falls to zero under every reinforcement requirement (minimum two or three successful digital neighbours, minimum two or three better-performing neighbours of any kind) and under the all-neighbour denominator. This reflects the small denominator at degree four, not a property of the rule: Appendix K (Experiments K and L) shows that at realistic, human-scale degree the same rule is genuine reinforcement-based complex contagion and survives the strict ≥2 requirement under adequate ignition. The hazard and frequency-based nulls remain at zero throughout.

## Appendix H: Cascade-metric sensitivity (Experiment H)

Experiment H reclassifies the existing run trajectories (no resimulation) over the grid of final-adoption cutoff {0.50, 0.70}, transition window W {30, 45, 60}, and minimum slope s_min {0.01, 0.015, 0.02, 0.025, 0.03}, with Wilson 95% confidence intervals; the full grid is in the replication CSV. Summary: the hazard and frequency-based rules are classified as cascades nowhere in the grid (probability 0.00). Performance-conditioned imitation in the Experiment A cell is 0.99 at the baseline (final 0.5, W = 45, s_min = 0.02), rises to 1.00 at the loosest setting (final 0.5, W = 60, s_min = 0.01), and remains 0.60 at the strictest (final 0.7, W = 30, s_min = 0.03). Intermediate transition cells shift in absolute probability with W and s_min but no headline ordering reverses, so the conclusions do not depend on the specific baseline values; W = 45 and s_min = 0.02 are retained as a middle-of-grid convention.

## Appendix I: Replacement-dynamics ablation (Experiment I)

Experiment I ablates the two replacement-entry rules (Section 3.9) across mechanism, transition, and placement cells (R = 250). The audit established that the production code for Experiments A–F set replacement entrants to digital with probability equal to the seed fraction (bernoulli_seed_probability); the model baseline is that replacement entrants are traditional (traditional_only). Cascade probabilities are essentially unchanged between the two rules (Table 8, Section 6.5): for example 1.00 vs 0.99 (Experiment A performance-conditioned), 0.00 vs 0.00 (hazard), 0.56 vs 0.58 ($\beta = 0.4$ mixture $\pi = 5\%$), 0.44 vs 0.52 (random/random placement), and 1.00 vs 1.00 (hub/bridge placement). The main conclusions are therefore robust to the replacement-entry rule; each results table notes which rule it used.

## Appendix: Parameter-space maps (Experiment J)

Experiment J maps cascade probability over three parameter planes (R = 100 per cell; production replacement rule). The $\beta \times$ misfit-share map appears in the main text (Figure 10); the $\bar{\theta} \times$ seed-fraction and digital-premium × variance maps are below. Together they show the mechanism operating over a bounded region: it requires sufficient network reach, a sizeable digital premium (stronger at lower variance), and enough well-placed seeds or low-threshold pioneers.

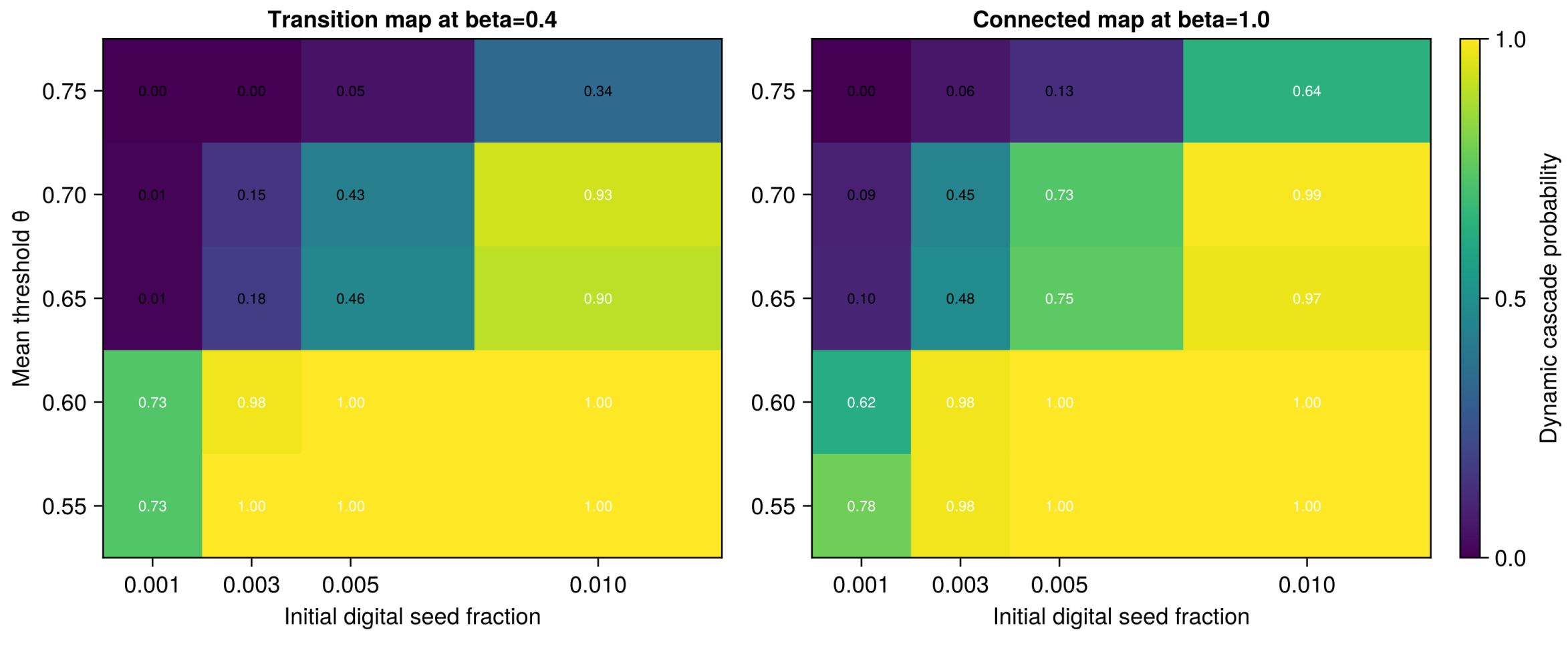

**Figure J1.** *Experiment J — cascade probability over mean threshold $\bar{\vartheta}$ and seed fraction (R = 100; production replacement rule).*

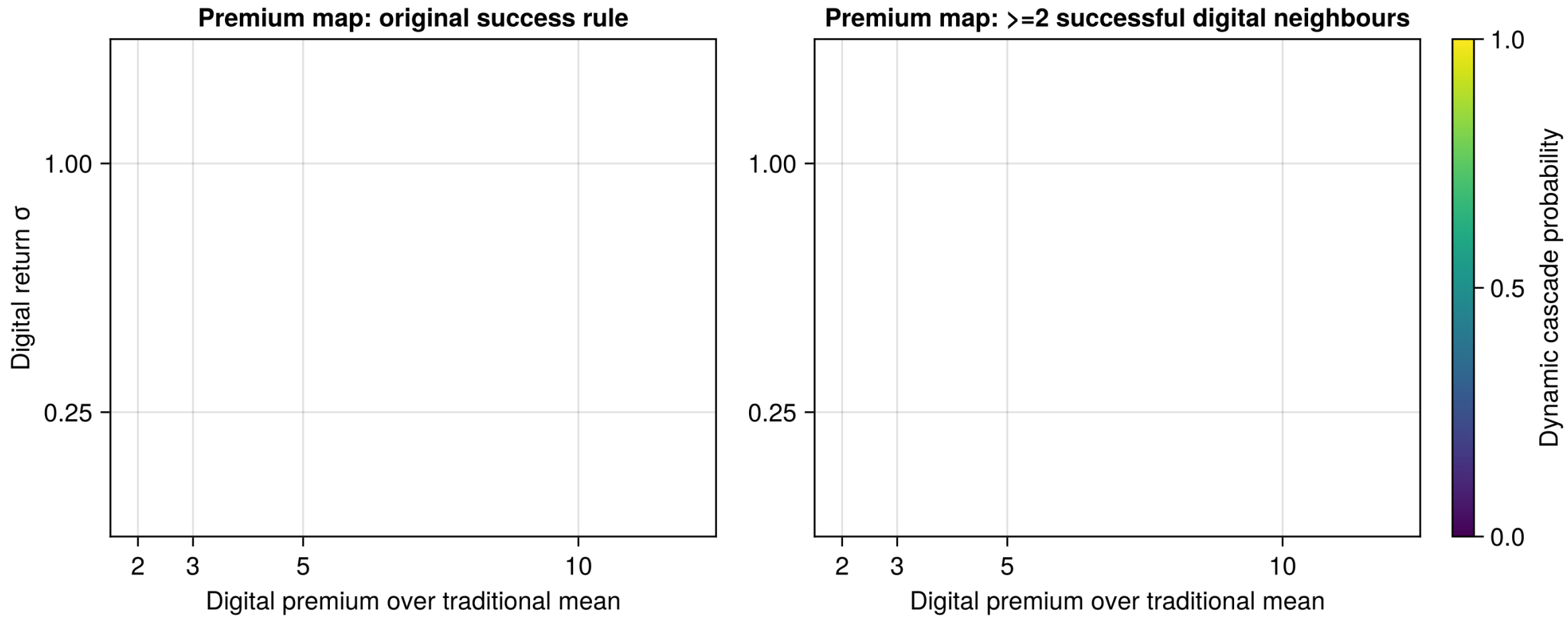

**Figure J2.** *Experiment J — cascade probability over digital premium and digital return variance (R = 100; production replacement rule).*

## Appendix K: Realistic-degree complex contagion (Experiments K and L)

Experiments K and L re-run the model at realistic, degree-controlled observation neighbourhoods (k = 10–150), recalibrating the phase boundary at each degree, and add a per-adoption diagnostic recording how many successful (better-performing, digital) neighbours a firm observes when it switches. Three results, against pre-registered criteria, recover the complex-contagion reading at realistic scale.

**Reinforcement (C2).** The median number of successful digital exemplars at adoption is ≈ 9 at k = 20, ≈ 25 at k = 50, and ≈ 38 at k = 100, with exposure share ≈ 1.0 — adoption follows broad local reinforcement, not a single exposure.

**Ignition (C3).** With an adequate, preferentially clustered seed, the explicit ≥2 successful-digital rule ignites and cascades (Table K1): e.g. at k = 50 the ≥2 rule reaches ≈ 0.53–0.68 and the ≥3 rule ≈ 0.37–0.43 once the seed exceeds ≈ 0.5–2% of firms, whereas the original fractional rule ignites from a 0.1% seed. Reinforcement-based contagion thus requires a critical seed mass to start — the Centola–Watts signature — rather than failing outright.

**Table K1.** *Experiment L — minimum seed for the strict rules to ignite (cascade probability with CI excluding zero; R per cell as logged).*

| Rule / degree | Min. seed | Best placement (cascade prob.) | $t_{50}$ (median) |
|---|---|---|---|
| Original fractional, k = 20 | 0.1% | hub/bridge 1.00 | 25 |
| ≥2 successful digital, k = 20 | 2% | hub/bridge 0.53 | 47 |
| ≥2 successful digital, k = 50 | 0.5% | hub/bridge 0.68 | 48 |
| ≥3 successful digital, k = 50 | 1–2% | clustered 0.43 | 53 |

**Bounded scale (C4).** Cascade probability is non-monotonic in degree: 1.0 at k = 10/15/30/50, then 0.0 at k = 100/150/200 (Table K2). The mechanism operates within human-scale neighbourhoods and collapses past roughly k ≈ 50–100, consistent with bounded cognition (Gigerenzer) and the subdivision of over-large clusters (Dunbar).

**Table K2.** *Experiment L — cascade probability versus observation degree (clustered seed, original rule).*

| k | 10 | 15 | 30 | 50 | 100 | 150 |
|---|---|---|---|---|---|---|
| Cascade prob. | 1.0 | 1.0 | 1.0 | 1.0 | 0.0 | 0.0 |

Per the pre-registered criteria, the realistic-degree model satisfies mechanism specificity, reinforcement, ignition under adequate seeding, and the bounded-scale prediction. Full grids, calibration-by-degree, and the exposure-at-adoption distributions are in the replication package.

## Appendix J: Reproducibility package

The model is implemented in Julia 1.10.4 using Agents.jl 6.2.10 and Graphs.jl 1.13.1, with CSV.jl 0.10.16, DataFrames.jl 1.8.2, and CairoMakie 0.13.10. Every experiment writes per-run, per-period, and summary CSVs plus a metadata JSON recording the master seed, per-run seeds, settings, and package versions. Scripts cover calibration, Experiments A–L, the topology regression, the realistic-degree and ignition experiments, and the stylized-facts check; figures are regenerated from the CSVs. A quick-mode driver reproduces the full set in minutes; a full-mode flag produces the paper-scale runs reported here. The repository contains the code, the full ODD, the parameter tables, the implementation audit, and replication instructions, under an MIT licence (code) and CC BY 4.0 (data and figures). The repository URL and archived DOI are withheld here for anonymous review and have been provided to the editor.

### J.1 Pseudocode

The complete model logic is summarised below. One period executes GROW, then EXIT-AND-REPLACE, then a synchronous ADOPT step; the adoption decision differs only by rule. Adoption is absorbing.

```
INITIALISE(N, family, k, p_seed, treatment, params):
    G ← build_network(family, N, k)          # degree-conserving Watts–Strogatz, BA, or ER
    θ[1..N] ← draw_thresholds(treatment, params)   # homogeneous | lower_mean |
                                                   # higher_variance | mixture(π, θ_low)
    seed_set ← place_seeds(G, n_seed, strategy)    # random | high_degree | bridge | clustered
    for each firm i on a node of G:
        digital[i]   ← (i ∈ seed_set)
        valuation[i] ← v0;   age[i] ← 0

STEP(model):                                 # run once per period, for t = 1..T
    # 1) GROW
    for each firm i:
        (μ, σ) ← digital[i] ? (1.5, 1.0) : (0.15, 0.10)
        g ← TruncatedNormal(μ, σ, lower = −0.99)
        valuation[i] ← valuation[i] × (1 + g);   age[i] ← age[i] + 1
    # 2) EXIT AND REPLACE  (N held constant)
    for each firm i with valuation[i] < 1:
        valuation[i] ← v0;  age[i] ← 0
        digital[i] ← Bernoulli(p_seed);  θ[i] ← redraw(treatment, params)
    # 3) ADOPT  (synchronous: decide on a frozen snapshot, then commit)
        to_adopt ← ∅
        for each firm i with digital[i] = false:
            if ADOPT?(i):  to_adopt ← to_adopt ∪ {i}
        for each i ∈ to_adopt:  digital[i] ← true     # absorbing

ADOPT?(i):                                   # returns true / false
    if rule = HAZARD:                              # exogenous null
        return rand() < λ
    if rule = FREQUENCY:                           # bandwagon / simple contagion
        Nb ← neighbours(i)
        if Nb = ∅: return false
        ϕ ← |{ j ∈ Nb : digital[j] }| / |Nb|
        return ϕ ≥ θ[i]
    if rule = SUCCESS:                             # imitate-the-successful / complex
        B ← { j ∈ neighbours(i) : valuation[j] > valuation[i] }
        if B = ∅: return false                     # operative candidate condition
        ϕ ← |{ j ∈ B : digital[j] }| / |B|
        return ϕ ≥ θ[i]
```

```
OUTCOMES(run):
    A(t) ← mean_i digital[i]   at each period t
    T10, T50, T90 ← first t with A(t) ≥ 0.1, 0.5, 0.9
    cascade ← (A(T) ≥ 0.5)  ∧  (T50 − T10 ≤ W)  ∧  (max_t ΔA(t) ≥ s_min)
    # baseline W = 45 periods, s_min = 0.02 per period
```